\documentclass{aastex63}

\usepackage{comment}
\usepackage{amsmath}
\usepackage{lineno}
\received{}
\revised{}
\accepted{}
\submitjournal{ApJ}

\shorttitle{}
\usepackage[utf8]{inputenc}
\usepackage{bm}
\usepackage[caption=false]{subfig}
\begin{document}

\title{Harmonic radio emission in randomly inhomogeneous plasma}

\correspondingauthor{Anna Tkachenko}
\email{anna.tkachenko@cnrs-orleans.fr}

\author[0000-0003-3705-9690]{Anna Tkachenko}
\affiliation{LPC2E, CNRS and University of Orléans, 
3A avenue de la Recherche Scientifique, Orléans, France}

\author[0000-0002-6809-6219]{Vladimir Krasnoselskikh}
\affiliation{LPC2E, CNRS and University of Orléans, 
3A avenue de la Recherche Scientifique, Orléans, France}
\affiliation{SSL, University of California at Berkeley,
\& Gauss Way, 94720, Berkeley, CA, USA}

\author[0000-0001-8307-781X]{Andrii Voshchepynets}
\affiliation{The Swedish Institute of Space Physics (IRF), Rymdcampus1, Kiruna, Sweden}
\affiliation{IRAP, 9 avenue du Colonel Roche, Toulouse, France}

%\collaboration{1}{(AAS Journals Data Scientists collaboration)}

%% Note that the \and command from previous versions of AASTeX is now
%% depreciated in this version as it is no longer necessary. AASTeX 
%% automatically takes care of all commas and "and"s between authors names.

%% AASTeX 6.3 has the new \collaboration and \nocollaboration commands to
%% provide the collaboration status of a group of authors. These commands 
%% can be used either before or after the list of corresponding authors. The
%% argument for \collaboration is the collaboration identifier. Authors are
%% encouraged to surround collaboration identifiers with ()s. The 
%% \nocollaboration command takes no argument and exists to indicate that
%% the nearby authors are not part of surrounding collaborations.

%% Mark off the abstract in the ``abstract'' environment. 
\begin{abstract}

%Type III radio bursts are 
In present paper, we describe a theoretical model of generation of harmonic emissions of type III solar radio bursts. The goal of our study is to fully take into account the most efficient physical processes participating in generation of harmonic electromagnetic emission via nonlinear coupling of Langmuir waves in randomly inhomogeneous plasma of solar wind ($l+l^{'} \rightarrow t$). We revisit the conventional mechanism of coalescence of primarily generated and back-scattered Langmuir waves in quasihomogeneous plasma. Additionally, we propose and investigate another mechanism that generates the harmonic emission only in a strongly inhomogeneous plasma: the nonlinear coupling of incident and reflected Langmuir waves inside localized regions with enhanced plasma density (clumps), in the close vicinity of the reflection point. Both mechanisms imply the presence of strong density fluctuations in plasma. We use the results of a probabilistic model of beam-plasma interaction and evaluate the efficiency of energy transfer from Langmuir waves to harmonic emission. We infer that harmonic emissions from a quasihomogeneous plasma are significantly more intense than found in previous studies. The efficiency of Langmuir waves conversion into electromagnetic harmonic emission is expected to be higher at large heliospheric distances for the mechanism operating in quasihomogeneous plasma, and at small heliocentric distances - for the one operating in inhomogeneous. The evaluation of emission intensity in quasihomogeneous plasma may also be applied for type II solar radio bursts. The radiation pattern in both cases is quadrupolar, and we show that emission from density clumps may efficiently contribute to the visibility of harmonic radio emission.        

\end{abstract}

%% Keywords should appear after the \end{abstract} command. 
%% See the online documentation for the full list of available subject
%% keywords and the rules for their use.

%% From the front matter, we move on to the body of the paper.
%% Sections are demarcated by \section and \subsection, respectively.
%% Observe the use of the LaTeX \label
%% command after the \subsection to give a symbolic KEY to the
%% subsection for cross-referencing in a \ref command.
%% You can use LaTeX's \ref and \label commands to keep track of
%% cross-references to sections, equations, tables, and figures.
%% That way, if you change the order of any elements, LaTeX will
%% automatically renumber them.
%%
%% We recommend that authors also use the natbib \citep
%% and \citet commands to identify citations.  The citations are
%% tied to the reference list via symbolic KEYs. The KEY corresponds
%% to the KEY in the \bibitem in the reference list below. 

\section{Introduction} \label{sec:intro}

Radio emissions in the inner heliosphere, associated with extreme space weather events, such as solar flares, are generated via a plasma emission mechanism first suggested by \cite{ginzburg1958possible}. Generally, it can be described as consisting of two steps: a beam of electrons, accelerated at reconnection sites of solar flares or excited by coronal mass ejections 
(CMEs) driven shock waves, is propagating from the Sun along the opened magnetic field lines. On its way, it interacts with the ambient plasma, generating Langmuir waves (with frequencies close to the local plasma frequency $\omega_{p_e}$) via bump-on-tail instability. In turn, these waves may transfer part of their energy into electromagnetic radio emission at the fundamental frequency (close to the local plasma frequency) and it's second harmonic (about twice the plasma frequency). This mechanism is widely recognized as responsible for the generation of type II and type III solar radio bursts. Type IIIs are normally associated with solar flares and energetic electrons beams with typical velocities $0.1-0.5c$, whereas type IIs are attributed to CMEs and less energetic electron beams. 

Most type III radio bursts exhibit harmonic structure. Up to decametric wavelengths, the fundamental and second harmonic (further referred to as simply \textit{harmonic}) components can often be distinguished, when occurring simultaneously. This wavelength range is supposed to be associated with the coronal plasma. Within the interplanetary medium, at larger wavelengths, it is almost impossible to separate one component from another, except for rare cases when electron beams are observed \textit{in situ} by spacecrafts \citep{kellogg1980fundamental, dulk1998electron}.  There is a big observational base of radio bursts of type III, made by spacecrafts (e.g., Wind, STEREO, PSP), as well as by ground-based radio telescopes (e.g., LOFAR, Nançay), which covers various frequency ranges. Type IIIs, due to their mechanism of generation, are a powerful tool for diagnostics of coronal and solar wind plasma, as well as for tracking energetic electron beams \citep{mann2018tracking}.  However, there still remain open questions associated with type III radio bursts, and many of them are related to the role played by background density fluctuations of the solar wind and corona \citep{robinson1998fundamental, reid2014review, chen2018fine}.

Random density fluctuations within the solar wind and solar corona are a well-known feature of the heliospheric plasma  \citep{neugebauer1975enhancement, goldstein1995magnetohydrodynamic, shaikh2010turbulent, chen2012density}.
The density spectrum typically exhibits a two-knee power law within the frequency domain $10^{-3}\div10^{2}$ Hz, with a breaking frequency around 0.6 Hz. Part of the spectrum below the break follows quite well the Kolmogorov power law, while the spectral index above the break depends on the conditions in the solar wind. Reported values of the spectral index above the break are typically in the range [-0.91, -0.38] \citep{celnikier1987aspects}. However there are reports (e.g., by \cite{kellogg2005rapid}) that show that the spectrum of density fluctuations can be well described by a single power law with spectral index
-1.37. Recent observations as well as \textit{in situ} measurements by Parker Solar Probe (PSP) spacecraft confirm, that the level of density fluctuations in the solar wind can go up to seven percent of the average background density at $\sim$ 36 $R_\odot$, and the analysis based on comparison of Monte Carlo simulations with PSP observations of decay times of the type III bursts predicts a growth of the level of density fluctuations towards the Sun, reaching up to twenty percent in the high corona \citep{krupar2020density}. These turbulent structures strongly affect the propagation and observed properties of radio emissions in coronal and solar wind plasmas \citep{kontar2019anisotropic}. At the same time, the presence of density fluctuations has a significant impact on the generation of Langmuir waves \citep{reid2010solar, krafft2013interaction, bian2014resonance, voshchepynets2015probabilistic}, which are later converted in electromagnetic (EM) emission. As it was shown by \cite{voshchepynets2015probabilistic2}, part of the density  spectrum in the range  $10^{-2}\div10^{1}$ Hz is of particular importance for the bump-on-tail instability under typical conditions of the solar wind plasma.  

According to the conventional plasma emission mechanism, the EM emission at a fundamental frequency is generated due to the Rayleigh scattering of Langmuir waves by plasma thermal ions, whereas the harmonic emission is the result of the Raman scattering of Langmuir waves \citep{ginzburg1958possible}. However, some of the observed properties of the type III radio emissions were not fully explained by the plasma emission mechanism. This has led to numerous revisions of the initial theory (e.g., see the review by \cite{reid2014review}). Several mechanisms were proposed, aiming to cover the properties of the fundamental emissions. Among them - the nonlinear wave-wave interaction of Langmuir, ion sound, and
EM waves: $ l\pm s\rightarrow t$ (e.g., see \citep{gurnett1978ion, melrose1987plasma}). The most important question concerning the efficiency of such process is the presence of ion sound waves with wavevectors $\mathbf{k}_{s} \approx \pm \mathbf{k}_{l}$ required to satisfy the resonant condition $\mathbf{k}_{l} \pm \mathbf{k}_{s} \rightarrow \mathbf{k}_{t}$, where $k_l = \omega_{p_e}/v_b \gg k_t=\omega_{p_e}/c$ and $v_b$ is the characteristic velocity of the beam. It means that either the broad spectrum of ion sound waves should contain the required waves or they should be generated by the decay instability, otherwise there is a high probability that some other process is responsible for generation of the EM emission at the fundamental frequency \citep{melrose1987plasma}. The ion sound wave, generated by the decay instability $l \rightarrow l^{'}+s$ in general cannot be directly involved in generation of the electromagnetic emission since the secondary Langmuir wave has the wavevector almost opposite to the wavevector of the primary wave, the wavevector of the sound wave is typically too large $\mathbf{k}_{s} \approx -2 \mathbf{k}_{l}$. Another mechanism that may explain the EM emissions at the fundamental frequency is the linear mode conversion (LMC) of Langmuir waves directly into EM waves in presence of an increasing density gradient. The importance of the encounters of Langmuir waves with the regions with higher density for the generation of EM waves was pointed out by a number of authors (e.g., see \citep{hinkel1992analytic} and references therein) and extensively studied by \cite{mjolhus1983linear, mjolhus1990linear, kim2007extraordinary, kim2008mode, kim2009waves, kim2013linear, schleyer2013linear, schleyer2014linear} for the case of a magnetized plasma. It was shown by \cite{krasnoselskikh2019efficiency} that even a simple reflection of Langmuir waves on a density inhomogeneity can result in an efficient generation of fundamental EM emission. It raises an important question about the possible role of similar localized regions where the reflection process can take place for the generation of the harmonic EM emission. 

For the harmonic emission with the frequency about $2\omega_{p_e}$, the wavevector of EM wave is $k_{t} \approx \sqrt{3}%
\omega_{p_e}/c$, while the primary Langmuir wave's wavevector is $k_{l} \approx \omega_{p_e}/v_{b}$. Thus the wavevector of the electromagnetic wave is much smaller than the wavevector of the Langmuir wave: $k_{t} \ll k_{l}$. It leads to the conclusion that the two coalescing Langmuir waves that produce an EM wave at $2 \omega_{p_e}$ should be almost antiparallel to fulfill the momentum conservation $\mathbf{k}_l + \mathbf{k}_{l'} = \mathbf{k}_t$. In the original work by \cite{ginzburg1958possible}, harmonic emission was attributed to the induced scattering of beam-driven Langmuir waves on thermal ions and subsequent coalescence of the forward-moving and scattered waves. However, a few decades later \cite{melrose1980emission, melrose1980plasma} has argued that ion scattering is not efficient enough to be consistent with the brightness temperatures of type IIIs, observed in corona. This has lead to the investigation of the role of ion sound waves in producing the back-scattered Langmuir waves via the electrostatic decay process $l \rightarrow l^{'}+s$ (e.g., see \citep{cairns1987second} and the references therein). Based on this idea, \cite{willes1996second} have derived the analytical solutions to describe the $l+l^{'} \rightarrow t$ process for a broad class of Langmuir waves spectra, including the case of almost anti-parallel waves (a head-on approximation). The ion sound waves, playing role in such process, should be generated in highly non-isothermal plasma with the electron temperature $T_e$ much larger than the ion temperature $T_i$ ($T_e \gg T_i$) \citep{chen1984introduction}. If this condition is not satisfied, the ion sound waves are quickly damped by resonant interactions with ions by means of Landau damping. Observations indicate, that in solar wind typically $T_e \sim T_i$ (e.g., \citep{lin1986evidence}), which leads to a conclusion that the electrostatic decay might be insufficient to account for the observed properties of harmonic type III emissions. 

Among the other mechanisms, suggested to explain harmonic emissions in plasma, there is the radiation by localized bunches of Langmuir waves \citep{galeev1976strong, brejzman1978electromagnetic, papadopoulos1978solitons, goldman1980radiation, ergun2008eigenmode, malaspina2012antenna}, which considers a radiation by nonlinear currents at twice the plasma frequency (antenna-type radiation). Some of these works imply the existence of a strong turbulence. However, in presence of strong density fluctuations, this approach becomes invalid. \cite{ergun2008eigenmode} have suggested another mechanism, based on the idea that a significant fraction of the Langmuir waves are localized as eigenmodes in solar wind density cavities. To enable this mechanism, density irregularities should have the form of density holes to capture Langmuir waves inside them. This is not always the case, since it requires the presence of localized density depressions. More common topological features are density clumps with a varying steepness of their density gradient. The theory of the harmonic emission in inhomogeneous plasma has already been partially considered, for example by \cite{erokhin1974theory} in one-dimensional case. Our goal is to revisit the theory of the harmonic emission by taking into account the density fluctuations within the solar wind and corona. In this case, the back-scattered Langmuir waves, required for the generation of the EM emission, are produced by the reflection of the forward moving Langmuir waves from density fluctuations. We consider two different cases of the harmonic EM emission generation: (1) the emission that is produced in the vicinity of the reflection point inside the clump and (2) the emission that is produced far from the clump, where the plasma can be considered as quasihomogeneous.

Several studies have also investigated the generation of third and higher harmonics, applicable to type II and III radio bursts \citep{brejzman1978electromagnetic, cairns1987third}.

A theoretical model of beam-plasma interaction
in a plasma with random density fluctuations developed by
\cite{voshchepynets2015probabilistic, voshchepynets2015probabilistic2}, shows that the reflection process is indeed very important
when the level of density fluctuations becomes large enough to overcome the effect of linear dispersion \citep{kellogg1999langmuir}, namely when 
\begin{equation}
\frac{\langle \Delta n \rangle}{n_0} > 3 {k_l^2}\lambda _{D}^{2}  = 3\frac{k_B T_{e}}{E_{b}}, 
\end{equation}
where $\langle \Delta n\rangle$ is the average level of density fluctuations, $n_0$ is the average background plasma density, $\lambda_D$ is the Debye length, $k_B$ is the Boltzmann constant, and $E_{b}$ is the characteristic kinetic
energy of the beam. It was also shown that for the aforementioned conditions the portion of wave energy carried by reflected waves approaches $50\%$ of the total wave energy, i.e., the energies of primary and reflected waves are almost equal. Thus, the process of reflection
of Langmuir waves on density fluctuations likely plays a crucial role for the
generation of EM emissions in quasihomogeneous plasma, even if it is treated as a
conventional coupling process that is not directly affected by density
fluctuations. On the other hand, in the close vicinity of the reflection points,
the electrostatic fields are known to be enhanced, which may lead to the
operation of the antenna type mechanism in these localized areas.

\begin{figure}[htbp]
\includegraphics[width = 1\textwidth]{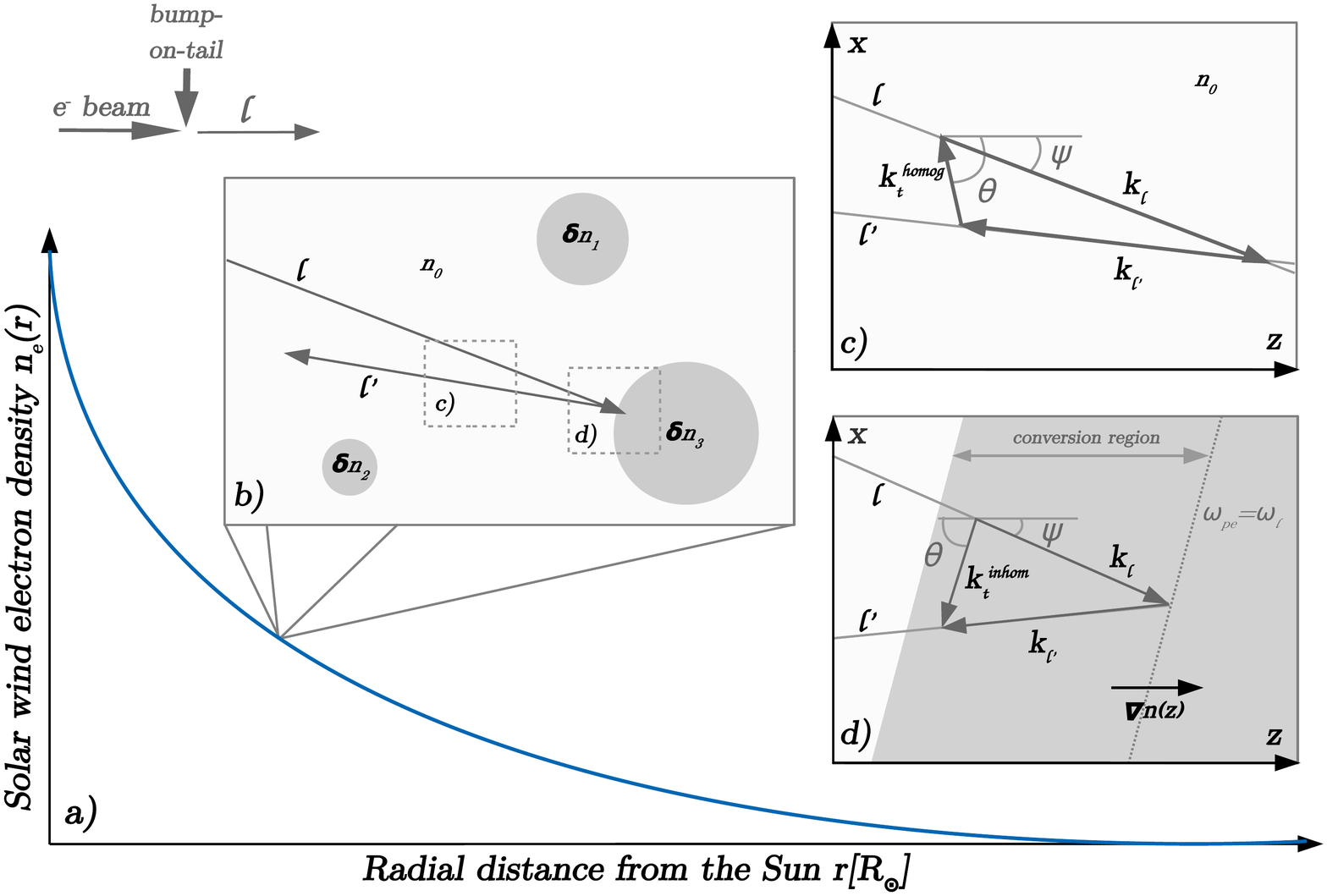}
\caption{Schematic illustration of the generation of harmonic EM emission in a randomly inhomogeneous plasma. Panel a) electron beam propagates from the Sun and generates a spectrum of Langmuir waves ($l$) with frequency $\omega_l$. The background electron plasma density decreases with the distance from the Sun (blue line). Panel b) Langmuir waves are reflected ($l'$) after an encounter with density clumps. Panel c) coalescence of two oppositely directed waves $\mathbf{k}_1$ and $\mathbf{k}_2$ from the spectra of Langmuir waves in quasihomogeneous plasma, the $z$-axis is directed along the electron beam direction, the $x$-axis is an arbitrary perpendicular direction.  Panel d) coalescence of the incident Langmuir wave with it's reflected part inside the conversion region, the $z$-axis is directed along the density gradient inside the clump,the $x$-axis is an arbitrary perpendicular direction. In panels c) and d)  $\psi$ is the acute angle between the wavevector of a forward moving (incident) Langmuir wave and the $z$-axis, and $\theta$ is the acute angle between the wavevector of harmonic EM wave and the $z$-axis. Generally, $z$-axes in these two different cases do not coincide. However, since the wavevectors of beam-generated Langmuir waves are usually highly aligned with the beam direction, and since we will consider coalescence in both regions under head-on approximation, we may assume that these axes are roughly equivalent.}
\label{fig:schem}
\end{figure}

In the present paper we study the process of generation of harmonic emission of type III solar radio bursts in a randomly inhomogeneous plasma. An electron beam resonantly generates a spectrum of Langmuir waves with a frequency $\omega_l$ very close to the local electron plasma frequency $\omega_{p_e}$ and with wavevectors highly aligned with the beam direction \citep{malaspina2008observations, krasnoselskikh2011determining}. As the average density of the background plasma decreases, random density fluctuations provide local density enhancements (clumps). Langmuir waves may encounter these clumps, and if the electron plasma frequency inside this structures reaches $\omega_l$, waves will be reflected in the opposite direction, forming a spectrum of backward moving (reflected) Langmuir waves (see Fig.\ref{fig:schem}b). The forward and backward moving Langmuir waves may then interact and produce harmonic EM emissions. We will formally distinguish two different regions of such interaction: (1) the quasihomogeneous plasma with average local electron density $n_0$ distant from the localized density perturbations (see Fig.\ref{fig:schem}c), and (2) the locally inhomogeneous plasma inside the density clumps, confined between the start of the positive density gradient and the reflection point, i.e., confined within the conversion region (see Fig.\ref{fig:schem}d). For the first case, we evaluate the process of nonlinear coupling of Langmuir waves assuming a mirror-type reflection and a Gaussian spectrum of forward moving and reflected Langmuir waves. The reflection process is taken into
account by means of the coefficient $P_{ref}$, characterizing the part of energy carried by 
reflected waves. The coupling process itself is described similarly to the one
in homogeneous plasma, assuming that it is not affected by the density fluctuations (Section \ref{sec: hom_coupl}). For the second case, we consider a coalescence of a single forward moving (incident) Langmuir wave with its reflected part inside the conversion region, in the close vicinity of the reflection point. We evaluate the electric fields of these Langmuir waves assuming a linear density gradient. This allows us to obtain the perturbations of the density and velocity of electrons, caused by the presence of Langmuir waves, and consequently we may evaluate the excited nonlinear currents at a frequency around $2\omega_{p_e}$ (Section \ref{sec:probl}). Next, we estimate the energy density of the EM emission produced by the aforementioned nonlinear currents, for such a single event of Langmuir wave reflection within a single density clump. In order to obtain the value of the energy density of EM emissions that corresponds to a full spectrum of Langmuir waves and to multiple reflections from density clumps with different amplitudes of density fluctuations and different characteristic scales of density gradient, we average our result over the relevant parameters (Section \ref{sec:emsingle}). In both cases, the energy density of EM harmonic emissions is expressed in terms of energy density of Langmuir waves. To obtain a quantitative evaluation and deduce scaling laws for the dependencies of EM wave intensity, we use the results of probabilistic model of beam-plasma interaction to describe the generation of Langmuir waves in a randomly inhomogeneous plasma (Section \ref{sec:elbeam}). Finally, we estimate the efficiency of conversion of Langmuir waves into harmonic EM emission for both considered regions of interaction (Section \ref{sec:effic}). The assumptions used for the evaluation of harmonic emission from a quasihomogeneous plasma also allow to apply the obtained results to type II solar radio bursts.

\section{Nonlinear coupling of Langmuir waves and generation of the EM \label{sec: hom_coupl}
emission in a quasihomogeneous plasma.}

Here we describe the process of generation of harmonic electromagnetic emission by nonlinear coupling of Langmuir waves in a quasihomogeneous plasma. We consider a coalescence of two oppositely directed Langmuir waves, one of which, $\mathbf{k}_1$, is aligned with the beam (it belongs to a spectrum of primary generated waves), and the other, $\mathbf{k}_2$, is supposed to be reflected by density fluctuations (it belongs to the spectrum of reflected waves) (see Fig.\ref{fig:schem}c). Even though density fluctuations play an important role here, providing reflected waves, we examine harmonic wave generation as happening in a quasihomogeneous plasma, i.e., far from localized density perturbations.         
The process of EM waves generation by the coupling of Langmuir waves is
described by the following set of equations \citep{tsytovich2012nonlinear}

\begin{equation}
\frac{dN_t(\mathbf{k}_{t})}{dt}=\int \int \frac{d^{3}\mathbf{k}_{1}d^{3}%
\mathbf{k}_{2}}{(2\pi )^{6}}w_{ll}^{t}(\mathbf{k_{1}},\mathbf{k}_{2},\mathbf{k}_{t})[N_l(%
\mathbf{k}_{1})N_l(\mathbf{k}_{2})-N_l(\mathbf{k}_{1})N_t(\mathbf{k}%
_{t})-N_l(\mathbf{k}_{2})N_t(\mathbf{k}_{t})], 
\label{eq: num_quanta}
\end{equation}

\begin{equation}
w_{ll}^{t}(\mathbf{k}_{1},\mathbf{k}_{2},\mathbf{k}_{t})=\frac{(2\pi
)^{6}e^{2}(k_{1}^{2}-k_{2}^{2})^{2}}{32\pi m_e^{2}\omega_{p_e}k_{t}^{2}}\frac{[%
\mathbf{k}_{1}\times \mathbf{k}_{2}]^{2}}{k_{1}^{2}k_{2}^{2}}\delta (\omega
_{t}-\omega _{1}-\omega _{2})\delta (\mathbf{k}_{t}-\mathbf{k}_{1}-\mathbf{k}%
_{2}), 
\end{equation}
where $N_l(\mathbf{k}_{1}), N_l(\mathbf{k}_{2})$ and $N_t(\mathbf{k}_{t})$ are the numbers of wave quanta of Langmuir and electromagnetic waves respectively, $\mathbf{k}%
_{1},\omega _{1},\mathbf{k}_{2},\omega _{2}$ are the wavevectors and frequencies of
Langmuir waves, $\mathbf{k}_{t},\omega _{t}$ are the wavevector and frequency of the EM
wave, $e$ and $m_e$ are the electron charge and
mass. It is reasonable to suppose that the number of quanta of Langmuir waves is much larger than the number of quanta of EM waves, 
\begin{equation}
N_l(\mathbf{k}_{1}),N_l(\mathbf{k}_{2}) \gg N_t(\mathbf{k}_{t}), 
\end{equation}%
then the second and the third terms under the integral in Eq.(\ref{eq: num_quanta}) may be neglected. The number of quanta is related to the wave energy density via the following
formulas:

\begin{equation}
W_{l} (\mathbf{k}_{l}) =\omega_l N_l,\qquad W_{t} (\mathbf{k}_{t}) = \omega _{t}N_t.
\end{equation}%
We remind that Langmuir waves have a frequency $\omega_l$ very close to the local electron plasma frequency $\omega_{p_e}$ of the region where they were excited. The spectrum of Langmuir waves is formed by two processes: a direct excitation of Langmuir waves due to the bump-on-tail instability (thus it may be approximated as a Gaussian, centered at the resonant
wavevector k$_{b}$ with the width $\Delta k_b$), and by reflection of Langmuir waves, assuming this reflection to be of the mirror-type (i.e., also a Gaussian, centered at $-%
\mathbf{k}_{b}$ having the same width in the wavevector space). Here we note that a similar approximation was used by \cite{willes1996second}, but the results obtained here are very different, as will be discussed later. Thus, the spectrum of Langmuir waves has the form (for detail see Appendix \ref{app: 1}):

\begin{equation}
N_l \left( k_{\parallel },k_{\perp }\right) =\frac{1}{\pi ^{3/2}\Delta k_b^{3}}%
N\left[(1-P_{ref})\exp \left(-\frac{(k_{\parallel }-k_{b})^{2}+k_{\perp }^{2}}{\Delta k_b^{2}}\right)+P_{ref}\exp
\left(-\frac{(k_{\parallel }+k_{b})^{2}+k_{\perp }^{2}}{\Delta k_b^{2}}\right)\right]. 
\end{equation}

Here we choose the parallel direction to be the direction of beam propagation, $N$ is the total number of Langmuir waves quanta, $P_{ref}$ is the
reflection coefficient that defines the redistribution of wave energy between primary (forward moving) and reflected waves. In our calculations, we choose the $z$-axis to be directed along the direction of beam propagation,
and the $x$-axis is along the second component of the EM wave wavevector. After direct but slightly cumbersome calculations presented in Appendix \ref{app: 1}, one can
obtain the following equation for the energy density of harmonic EM emission in wavevector space, generated by the aforementioned Langmuir waves:%
\begin{equation}
\frac{dW_{t}^{homog}(\mathbf{k}_{t})}{dt}=\frac{P_{ref}(1-P_{ref})\omega_{p_e}}{192\pi ^{5}}%
\left( \frac{k_{b}}{\Delta k_b}\right) ^{4}\frac{\mathbf{k}_{t}^{2}}{k_{b}^{5}} \frac{%
W_l}{n_0 k_B T_e} W_l\sin ^{2} \psi \cos ^{2}\psi ,
\end{equation}
where superscript $'homog'$ indicates that we perform this estimation for a quasihomogeneous plasma, $W_l$ is the total energy density of Langmuir waves, and the angle $\psi $ is the angle between the vector $\mathbf{k}_{1}$ and the $z$-axis. Let us come back to the discussion of the fact that this result is very different from the one obtained by \cite{willes1996second} (see Eq.(29) of the aforementioned paper). \cite{willes1996second} have performed their calculation supposing that the following inequality is satisfied: 
\begin{equation}
\frac{2k_{r}k_{t}\sin \psi \sin \theta }{\Delta k^{2}}\gg1, 
\end{equation}%
and their approximate estimation is strongly dependent on this assumption. We carried out exact calculations without this assumption and the result obtained shows that the major input comes from the region in wavevector space where this inequality is not satisfied. It is worth noting here that there is a quite simple interpretation of this disagreement. The above assumption corresponds to neglecting currents parallel to the direction of propagation of Langmuir waves. On the other hand, the multiplier $\sin ^{2}\psi \cos ^{2}\psi $ indicates that the major directions of the emission comprise angles ${\pi }/{4}$ and ${3\pi }/{4}$ with the electron beam direction, unambiguously showing the quadrupolar character of the emission and the major input of these parallel electric currents. Integrating over $d^{3}\mathbf{k}_{t}$ one can evaluate 
\begin{equation}
\frac{dW_t^{homog}}{dt}= \frac{P_{ref}(1-P_{ref})\omega_{p_e}}{7200 \pi ^{4}}\left( \frac{k_{b}}{\Delta k_b}\right)
^{4}\frac{\mathbf{k}_{t}^{5}}{k_{b}^{5}} \frac{W_l}{n_0 k_B T_e} W_l.
\end{equation}
Observations show that the dynamics of the burst on its initial stage is
quite similar to the exponential growth from some noise level till the
maximum intensity is reached. Thus the Langmuir waves dynamics may be presented in the
form  
\begin{equation}
W_l=W_{noise}\exp (\gamma t ).
\end{equation}%
The characteristic growth factor until the instability saturation is supposed
to be of the order of $(\gamma t_{s} )=\Lambda =\ln (n\lambda _{D}^{3})$, thus 
\begin{equation}
\frac{W_l}{W_{noise}}=\exp \Lambda \text{, }
\end{equation}%
and the saturation occurs at a time $t_{s}= \Lambda / \gamma $. Consequently the equation for $%
W_t^{homog}$ may be solved as 
\begin{equation}
W_t^{homog}(t)= \frac{P_{ref}(1-P_{ref}) \omega_{p_e} }{7200\pi ^{4} \gamma }\left( \frac{k_{b}}{\Delta k_b}%
\right) ^{4}\frac{\mathbf{k}_{t}^{5}}{k_{b}^{5}} \frac{W_{l}^{2}}{n_0 k_B T_e}.
\end{equation}
Here $\gamma $ may be evaluated as $\gamma _{lin} / \Lambda $, where $\gamma _{lin}$ is the linear increment of the bump-on-tail instability in an inhomogeneous plasma with  random fluctuations. Parameters $W_{l}$ and $\gamma_{lin}$ will be estimated according to probabilistic model of beam-plasma interaction in a randomly inhomogeneous plasma in Section \ref{sec:elbeam}.  

\section{Description of the fields and currents in the vicinity of reflection points} \label{sec:probl}

Let us now consider the generation of the EM waves that may come from localized regions where the reflection of the Langmuir waves occurs (see Fig.\ref{fig:schem}d). In these regions, the field amplitudes and the corresponding currents are strongly affected by the inhomogeneity and here we present a model that intends to describe them precisely. We consider a Langmuir wave of frequency $\omega_l$ that encounters a density clump, formed due to random density fluctuations, inside which the density linearly increases towards the center. The assumption of a linear density gradient inside the clump is made for the sake of simplicity. As the wave enters the clump, the component of its wavevector in the direction of the density gradient decreases and in the point where the local plasma frequency reaches the frequency of a Langmuir wave, this wave will undergo a mirror-type reflection. After reflection, two waves, the incident $l$ and the reflected $l'$ will perturb electron trajectories, creating variations of the velocity and particle density that produce currents with frequency close to $2 \omega_{p_e}$ in the vicinity of the reflection point. These localized currents represent a source that can generate an EM wave $t$ with a frequency around $2 \omega_{p_e}$. Here we shall consider this process in more detail. We choose the density gradient inside the clump to be directed along the $z$-axis, while we assume that along the other two directions, $x$ and $y$, density variations are negligible, allowing us to reduce the number of dimensions and consider our problem only in the $(x,z)$ plane. The generation of the harmonic electromagnetic wave in this model is only possible for certain values of the angle of incidence $\psi$, since momentum conservation implies constraints on its value, determined by the ratio of the beam velocity and speed of light: $|\psi_{\max}| \approx {\sqrt{3}}{v_b} / 2{c}$ (see Appendix \ref{app: 3}).            

% \section{Electrostatic potential of a Langmuir wave} \label{sec:phiLang} %%%%%%%%Phi Lang

We begin our calculations with the system of equations for plasma oscillations  \citep{zakharov1972collapse}:

\begin{eqnarray}
\Delta \Phi_{l}^{inhom} (\bm{r},t) = 4 \pi e \delta n_l, \label{eq:zakharov1}\\
\frac{\partial }{\partial t} \delta n_l + \nabla ((n_0 + \delta n) 
\bm{v_e}) = 0, \\
\frac{\partial \bm{v_e}}{\partial t} = \frac{e}{m_e} \nabla \Phi_{l}^{inhom} (\bm{r},t)
- 3 v_{T}^2 \nabla \frac{\delta n_l}{n_0},
\label{eq:zakharov3}
\end{eqnarray}
where
\begin{displaymath}
n_e = n_0 + \delta n + \delta n_l.
\end{displaymath}
Here $\delta n$ is the given low-frequency plasma inhomogeneity, $\delta n_l$ is the
high frequency density variation caused by Langmuir oscillations, $\Phi_{l}^{inhom} (\bm{r},t)$ is the corresponding high-frequency part of the electrostatic potential. Here and further below, superscript $'inhom'$  refers to an inhomogeneous plasma inside a density clump. 

We assume that in a quasihomogeneous plasma outside the clump, the high-frequency part of electrostatic potential has the form of a plane wave: $\Phi_{l}^{homog} (\bm{r}, t) = \Phi_{0}^{homog} \exp(-i (\omega_{p_e} t + \eta) + i k_{l_x} x + i k_{l_{z0}} z)$, where $ \Phi_{0}^{homog}$ is the amplitude and $\eta$ is the phase difference between the wave in the homogeneous plasma and in the density clump. As we assume that all the parameters vary only along the $z$-axis, the solution for $\Phi_{l}^{inhom} (\bm{r}, t)$ can be written in a form $\Phi_{l}^{inhom} (\bm{r},t) = \Phi_{0}^{inhom} \phi (z) \exp(-i\omega_{p_e} t + i k_{l_x} x)$, where $\phi(z)$ is an unknown function of $z$. 

Without loss of generality, one can assume that the density gradient within the clump is linear, thus the density inhomogeneity profile has the form (density starts to increase at $z=0$): 

\begin{displaymath}
\frac{\delta n}{n_0} = \theta (z) \frac{z}{L},
\end{displaymath}
where $\theta (z)$ is the Heaviside step function, $L$ is a term, which can be interpreted as the characteristic scale of the density gradient. 
By introducing a new dimensionless variable
 
\begin{equation}
\tilde{z}=\left( \frac{k_l^2 L^{2}}{3 k_l^2 \lambda _{D}^{2}}\right) ^{1/3} \left(\frac{z}{L} - 3 k_l^2 \lambda_D^2 \cos^2 \psi \right),
\end{equation}
one can find from Eqs.(\ref{eq:zakharov1})-(\ref{eq:zakharov3}) a well-known Airy equation for the electrostatic potential:

\begin{equation}
\frac{d^2}{d\tilde{z}^2} \phi (\tilde{z}) - \tilde{z} \phi (\tilde{z}) = 0.
\end{equation}
It is convenient to present the solution making use of Hankel functions $H_{\nu}^{(n)}$ $(n = 1,2)$. They allow to easily separate the incident and reflected wave. Within the conversion region, the electrostatic potential is written in the following form :

\begin{equation}
\phi(\tilde{z}_{0} < \tilde{z} < 0) = \frac{1}{2} \sqrt{\frac{(-\tilde{z})}{3}} \left( e^{-i\pi/6} H^{(2)}_{1/3} \left(\frac{2}{3} (-\tilde{z})^{3/2} \right) + e^{i\pi/6} H^{(1)}_{1/3} \left(\frac{2%
}{3}(-\tilde{z})^{3/2} \right) \right) = \phi_{i} + \phi_{r}.
\end{equation}
Here $\tilde{z}_0 = - (3 k_l^2 \lambda_D^2)^{2/3} (k_l L)^{2/3} \cos^2 \psi$ denotes the beginning of the density gradient, the reflection of the Langmuir wave occurs at $\tilde{z} = 0$, the first term corresponds to the incident ($i$) wave and
the second term to the reflected ($r$) wave.
We evaluate the amplitude of electrostatic potential $\Phi_{l}^{inhom}$ (for the details see Appendix \ref{app: 2}) as a function of the electrostatic potential in a homogeneous plasma:

\begin{equation}
\Phi_{0}^{inhom} = 2 \sqrt{\pi} (3 k_l^2 \lambda_D^2 k_l L )^{1/6} \cos^{1/2} \psi \Phi_{0}^{homog}. 
\end{equation}

\begin{figure}[htbp]
\centering
\includegraphics[width=1\textwidth]{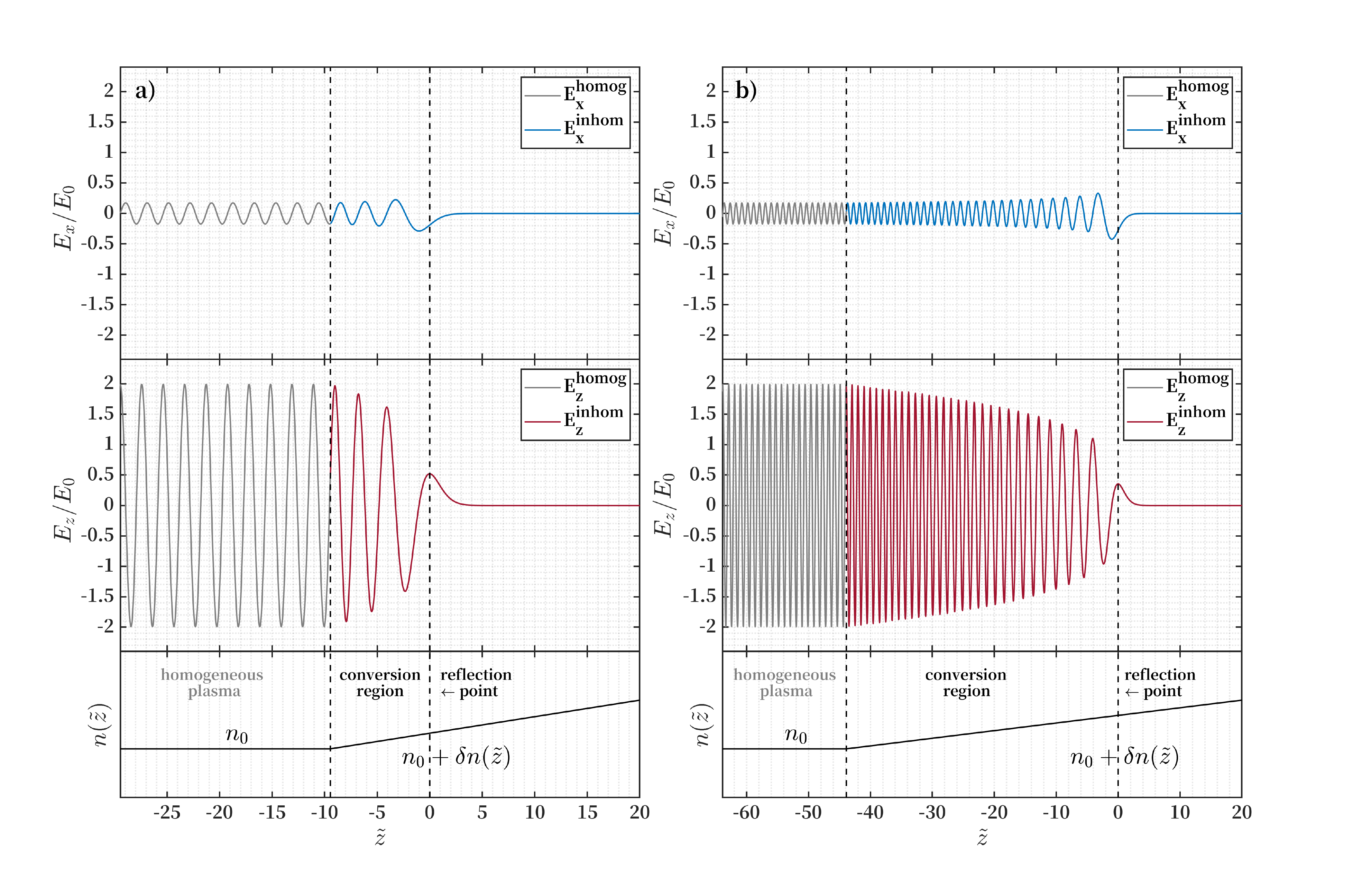} 
\caption{Absolute value of total electric field of incident and reflected Langmuir waves, normalized to the amplitude of the electric field in homogeneous plasma $E_0$, plotted versus a schematic illustration of the corresponding electron density profile. Parameters are: $T_e= 100$ eV, $v_b / c = 0.1$, $\psi = 5^o$. Panel a) $k_l L = 5\cdot 10^2$. Panel b) $k_l L = 5\cdot 10^3$.}
\label{fig: El_field}
\end{figure}

 %Behind the reflection point:
%\begin{equation}
%\phi(\tilde{z} > 0) = \frac{1}{\pi} \sqrt{\frac{\tilde{z}}{3}} K_{1/3} \left( \frac{2}{3} \tilde{z}^{3/2} \right).
%\end{equation}
Corresponding electric field components can be found as $\bm{E}_{l}^{inhom} = - \nabla \Phi_{l}^{inhom} (\bm{r},t)$ (Fig.\ref{fig: El_field}) which are expressed in terms of $E_0$ - the amplitude of the electric field in the homogeneous plasma, that can be found as $E_0 = k_l \Phi_0^{homog}$. 

\begin{figure}[htbp]
\centering
\includegraphics[width = 1\textwidth]{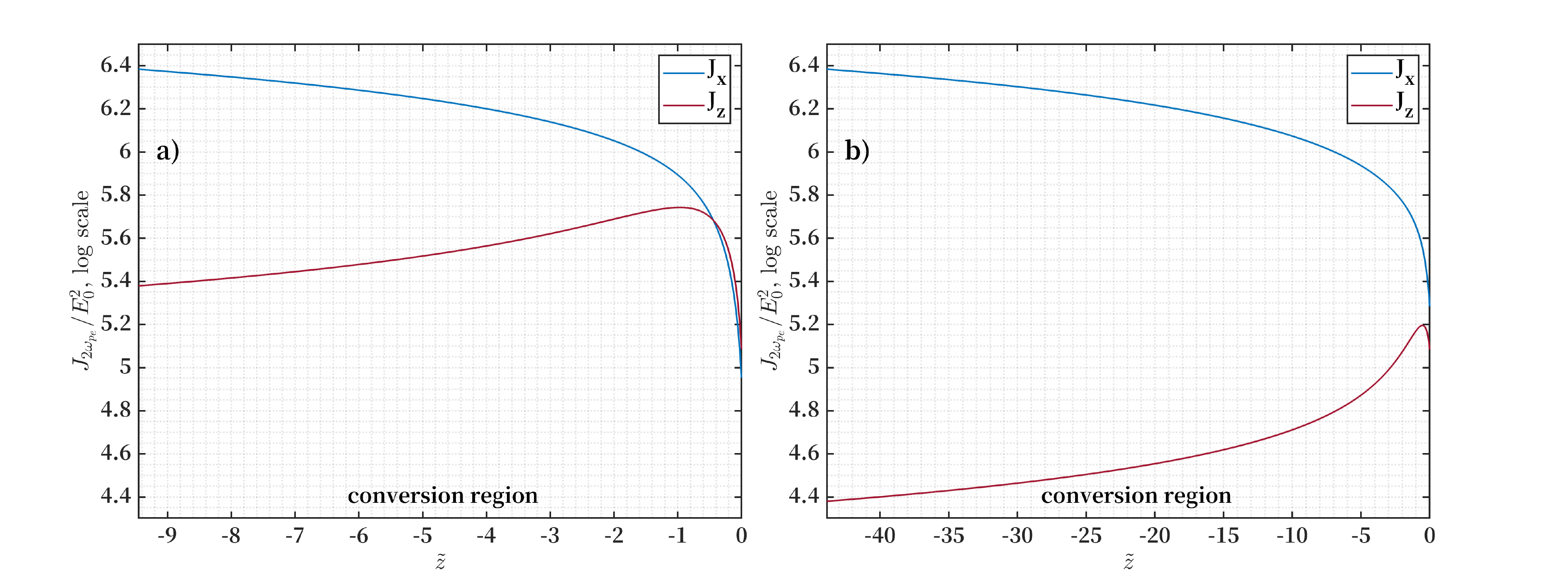}
\caption{Absolute value of harmonic current density, normalized to the square of the amplitude of electric field in homogeneous plasma $E_0$, inside the conversion region ($\tilde{z}_0 < \tilde{z} < 0$). Parameters are: $T_e = 100$ eV, $v_b / c = 0.1$, $\psi = 5^o$. Panel a) $k_l L = 5\cdot 10^2$. Panel b) $k_l L = 5\cdot 10^3$.}
\label{fig: curr}
\end{figure}

Having obtained the expressions for two separate components of the electric field, attributed to incident and reflected Langmuir waves, one can derive electron density and velocity perturbations excited in the plasma by each wave, using simple linear relations (see Appendix \ref{app: 2}). Thus, the nonlinear current resulting from superposition of these perturbations should be written in the following form:

\begin{equation} 
	\bm{J}_{2 \omega_{p_e}}(\bm{r}) = -e (\delta n_{l_{i}} (\bm{r}) \bm{\delta v}_{r} (\bm{r}) + \delta n_{l_{r}} (\bm{r}) \bm{\delta v}_{i} (\bm{r})). \label{eq: curr}
\end{equation}
This nonlinear current represents a localized source of generation of EM emission at about $2 \omega_{p_e}$ (see Fig.\ref{fig: curr}).

\section{Emission from localized density perturbations} \label{sec:emsingle} %%%%%%%%%% Emission

As the harmonic EM emission, generated in the corona or interplanetary medium, is mainly observed at distances much larger that its source size and wavelength, we will consider the EM field of this emission as such at a large distance from the source, which implies a decomposition \citep{landau2013course}:  
\begin{equation}
\bm{J}_{2 \omega_{p_e}}(\bm{r}, t - \frac{|\bm{R}-\bm{r}|}{c}) \approx \bm{J}_{2 \omega_{p_e}}(\bm{r}, t - \frac{R}{c} + \frac{\bm{r}\bm{n}}{c}) = \bm{J}_{2 \omega_{p_e}}(z) e^{-2 i \omega_{p_e} t + 2 i \omega_{p_e} R / c - 2 i \omega_{p_e} \bm{r}\bm{n} / c  + 2 i k_{l_x} x}.
\end{equation}
Here $\omega_{p_e} \bm{r}\bm{n} / c$ may be interpreted as the ratio of emission source size and the wavelength. The Liénard-Wiechert potential of the current is:
\begin{equation}
	\bm{A}_{2 \omega_{p_e}} (\bm{r}) e^{-2 i \omega_{p_e} t} = \frac{\sqrt{\epsilon}}{c}\int\frac{\bm{J}_{2 \omega_{p_e}}(\bm{r}, t - \frac{|\bm{R}-\bm{r}|}{c})}{|\bm{R}-\bm{r}|} d^3 r.
\end{equation}
It is well known that a dipolar emission is linearly proportional to the amplitude of oscillations of the center of mass of charged particles. For a system that consists only of electrons (we neglect ion motions for Langmuir waves) the center of mass may not undergo any displacement, thus the dipolar emission is absent \citep{landau2013course}. The absence of a dipolar component in the mechanism of generation of EM emission by localized wave packets was first pointed out by \cite{galeev1976strong}. So, we imply that quadrupole component of the emission is dominant. For the very same reason, the magnetic dipolar radiation is also absent. Thus we can use the decomposition
\begin{equation}
\frac{1}{|\bm{R}-\bm{r}|} \approx 4 \pi \sum_{l = 0}^\infty \sum_{m = -l}^l \frac{1}{2l+1} \frac{r^l}{R^{l+1}}Y^{m^*}_{l}(\theta_r, \phi_r) Y^m_l  (\theta_R, \phi_R),
\end{equation}
and keep only the terms corresponding to $l=2$ that account for the quadrupolar emission. The magnetic field of the EM wave is
\begin{equation}
\bm{H}_{t_{2 \omega_{p_e}}} = 2 i [\bm{k}_t \bm{A}_{2\omega_{p_e}}], 
\end{equation}
which can be rewritten as 
\begin{eqnarray}
| H_{t_{2 \omega_{p_e}}} | = H_y = 2i(-k_{t_x} A_{z_{2 \omega_{p_e}}} + k_{t_z} A_{x_{2 \omega_{p_e}}}).
\end{eqnarray}
We note that the term proportional to $A_{x_{2 \omega_{p_e}}}$ will be the most important as the current $J_{x_{2 \omega_{p_e}}}$ prevails over $J_{z_{2 \omega_{p_e}}}$ for larger angles (see Fig.\ref{fig: curr}), and for very small angles the product $k_{t_x} A_{z_{2\omega_{p_e}}} \sim \sin \psi A_{z_{2\omega_{p_e}}}$ vanishes. The radiant energy density of the emission is
\begin{equation}
W^{inhom}_t = \frac{|H_{t_{2 \omega_{p_e}}}|^2}{8 \pi}.
\end{equation}
An approximate analytical expression for $W^{inhom}_t$ from a single localized density clump is (for details see Appendix \ref{app: 3}):
\begin{equation} 
W_t^{inhom} = 4\cdot10^2 \pi \epsilon \frac{v_{T}^6}{v_b^6} \left({\frac{k_l^2 \lambda_D^2}{\delta n / n_0}}\right)^{2} \frac{\omega_{p_e^2} L^2}{c^2} \sin^2 \psi \frac{W_{l}}{n_0 k_B T_e} W_{l},
\end{equation}
where $v_T$ is a thermal velocity of electrons in the plasma, and $W_{l} = E_0^2 / 8 \pi$ is the energy density of Langmuir waves. This expression was derived under the approximation $3 k_l^2 \lambda_D^2 \ll \delta n / n_0$ and  $\sin \psi \ll 1$. 

On their way through the inhomogeneous solar wind, Langmuir waves can encounter density clumps of different size and magnitude. As it was shown earlier, both of these parameters can strongly affect the harmonic emission from inside the density clump. For instance, Fig.\ref{fig: El_field} demonstrates that with growth of $L$ (and decrease of the amplitude of density fluctuation $\delta n / n_0 \sim L^{-1}$) the conversion region increases, allowing the perpendicular current $J_{x_{2 \omega_{p_e}}}$ to grow significantly larger than $J_{z_{2 \omega_{p_e}}}$ for the major part of angles $\psi$. To take this into account, one can estimate statistically the averaged value of $W^{inhom}_t$. The size of the source region of type III bursts is typically much larger that the characteristic scale of density fluctuations within the solar wind \citep{reid2014review} and, thus, the number of encounters is large enough to justify averaging: 
\begin{equation}
\langle W_t^{inhom} \rangle_{\psi, \delta n, L} = P_{ref} \int P(\psi) P(\delta n) P(L) W_t^{inhom}(\psi, \delta n, L) \ d \psi\ d \delta n \ d L,
\end{equation}
here $P(\psi)$, $P(\delta n)$ and $P(L)$ are probability distribution functions (PDF) of the angle of incidence, amplitudes of the density fluctuations within the clumps and of the scales of gradients inside clumps, respectively. We suppose that angles $\psi$ are distributed uniformly between 0 and $\psi_{\max}$, the amplitudes $\delta n$ follow a normal distribution with zero mean and standard deviation $\langle \Delta n \rangle$, and scales follow the distribution adopted from \citep{krasnoselskikh2019efficiency} (see also Appendix \ref{app: 4}):
\begin{equation}
P_{L}\left( L\right) = \frac{1}{\sqrt{2\pi}} \frac{L_{sc}}{L^{2}}\exp{\left[-\frac{L_{sc}^{2}}{2L^{2}}\right]},
\end{equation}
where $L_{sc}$ is the characteristic scale of density gradients for normally distributed density fluctuations. It can be approximated as a function of the level of density fluctuations :
\begin{equation}
L_{sc} \approx 1.4 \cdot \left( \frac{\langle \Delta n \rangle}{n_0} \right)^{-1}, \ [km].    
\end{equation}
As it was shown by \cite{krasnoselskikh2019efficiency}, random density fluctuations, described this way in terms of $P(\delta n)$ and $P(L)$, reproduce the interval of density spectrum from $10^{-2}$ Hz to 530 Hz measured within the solar wind. The reason to choose this part of the spectrum is that the density fluctuations that may affect the beam-plasma interaction have the characteristic scales that on the one hand should be much larger than the wavelength of the Langmuir waves and, on the other hand, must be significantly smaller than the relaxation length of the beam-plasma interaction. The averaged value of the energy density of the harmonic EM emission is (for details see Appendix \ref{app: 4})
\begin{equation}
\langle W_t^{inhom} \rangle_{\psi, \delta n, L} = {25 \sqrt{6}} \epsilon P_{ref} \frac{v_{T}^3}{c^3} \frac{v_T^3}{v_b^3} \left({\frac{k_l^2 \lambda_D^2}{\langle \Delta n \rangle / n_0}}\right)^{2} \frac{\omega_{p_e^2} L_{sc}^2}{c^2} \frac{W_{l}}{n_0 k_B T_e} W_{l}.
\end{equation} 
To evaluate the efficiency of the generation mechanism, one should establish the relations between the parameters of the electron beam and the characteristics of spectra and temporal evolution of the Langmuir waves, generated via beam-plasma interaction. We present hereafter some results of the  probabilistic model of beam-plasma interaction in a plasma with random density fluctuations.

\section{Electron beam - plasma interaction in randomly inhomogeneous plasma} \label{sec:elbeam}

Before proceeding with EM harmonic emission, we should evaluate the wave energy
density of Langmuir waves. The problem of beam-plasma interaction may be
analyzed by means of quasilinear theory (QLT) which takes into account the
process of generation of Langmuir waves due to the bump-on-tail instability
of energetic electrons, and the following modification of the electron distribution
function that eventually leads to the formation of a plateau. An important condition for
a QLT description of this process consists in exact resonance between the wave and
the particle: in the one-dimensional case they interact only when the particle velocity is
exactly equal to the phase velocity of the wave \citep{vedenov1962quasi, drummond1964nonlinear}. 
There have been several reports exploring two-dimensional (e.g., by \cite{ziebell2008two, ziebell2011two}) and three-dimensional (e.g., by \cite{ harding2020electron}) quasilinear wave-particle interactions, that account for angular diffusion of wavevectors of Langmuir waves \citep{nishikawa1976relaxation, krasnoselskikh2007beam}. However, the one-dimensional approach remains justified here due to weak angular dispersion of the beam velocities and associated Langmuir waves reported at around 1 a.u. \citep{ergun1998wind, malaspina2008observations, krasnoselskikh2011determining}. For the case of a homogeneous plasma, the QLT predicts the formation of a plateau in the electron
velocity distribution function in the range of velocities from the beam
velocity to the thermal velocity of the plasma. The process of
"plateauing" is accompanied by a transformation of the free kinetic energy of
the electrons into the potential energy of Langmuir waves. \cite{sturrock1964type}
applied the QLT description of beam-plasma interaction to solar radio bursts.
The analysis of beam-plasma interaction under conditions relevant to
solar corona and solar wind resulted in so-called "Sturrock paradox": the
relaxation of the beam should have stopped after a very short distance, about 100 km. However, Langmuir waves and associated beams had been
observed up to the Earth orbit. Later, satellite measurements have shown
that such beams are observed in the solar wind even at distances of
about 5 a.u.

Recent studies \citep{kellogg2005rapid, Krucker_Oakley_Lin_2009,  Ratcliffe_Kontar_2012, Voshchepynets_Krasnoselskikh_2013} have
demonstrated that there is an important characteristic of the solar wind
that should be taken into account when analysing the aforementioned processes. The solar wind
is quite strongly inhomogeneous, filled with random density fluctuations ${\delta n}/{n_0}$
that may be quite intense, about several percent of the background plasma
density at 1 a.u. Under such conditions the phase velocity $V_{ph}$ of Langmuir waves varies, since the probability distribution of the phase velocity is determined by the probability distribution of the density
fluctuations due to relation 
\begin{equation}
\omega_{p_e}^{2}(n)=\omega_{p_e}^{2}(n_{0})(1+\frac{\delta n}{n_{0}})=\omega
^{2}(1-3v_{T}^{2}/V_{ph}^{2}),
\end{equation}
and consequently, the wave along its path resonantly interacts with electrons of
different velocities.
There are several complementary models that describe wave-particle
interaction in randomly inhomogeneous plasma. One is a Hamiltonian numerical
model where the background plasma is described by Zakharov equations \citep{krafft2013interaction, krafft2014hamiltonian,Volokitin_Krafft_2016,Volokitin_Krafft_2018,Volokitin_Krafft_2020}, and the beam and its interaction with
waves is modeled by a PIC code. In this model the system is periodic and is
chosen to be long enough to incorporate several modes of density
fluctuations. The second model is based on the use of the probability
distribution of the wave's phase velocity in the plasma with random density
fluctuations \citep{voshchepynets2015probabilistic}. The resonant wave-particle
interaction takes into account the wave interaction with particles having
different velocities. The probability distribution becomes a statistical
weighting function that results in a natural widening of the resonance
conditions.

\begin{figure}[htbp]
\includegraphics[width = 1\textwidth]{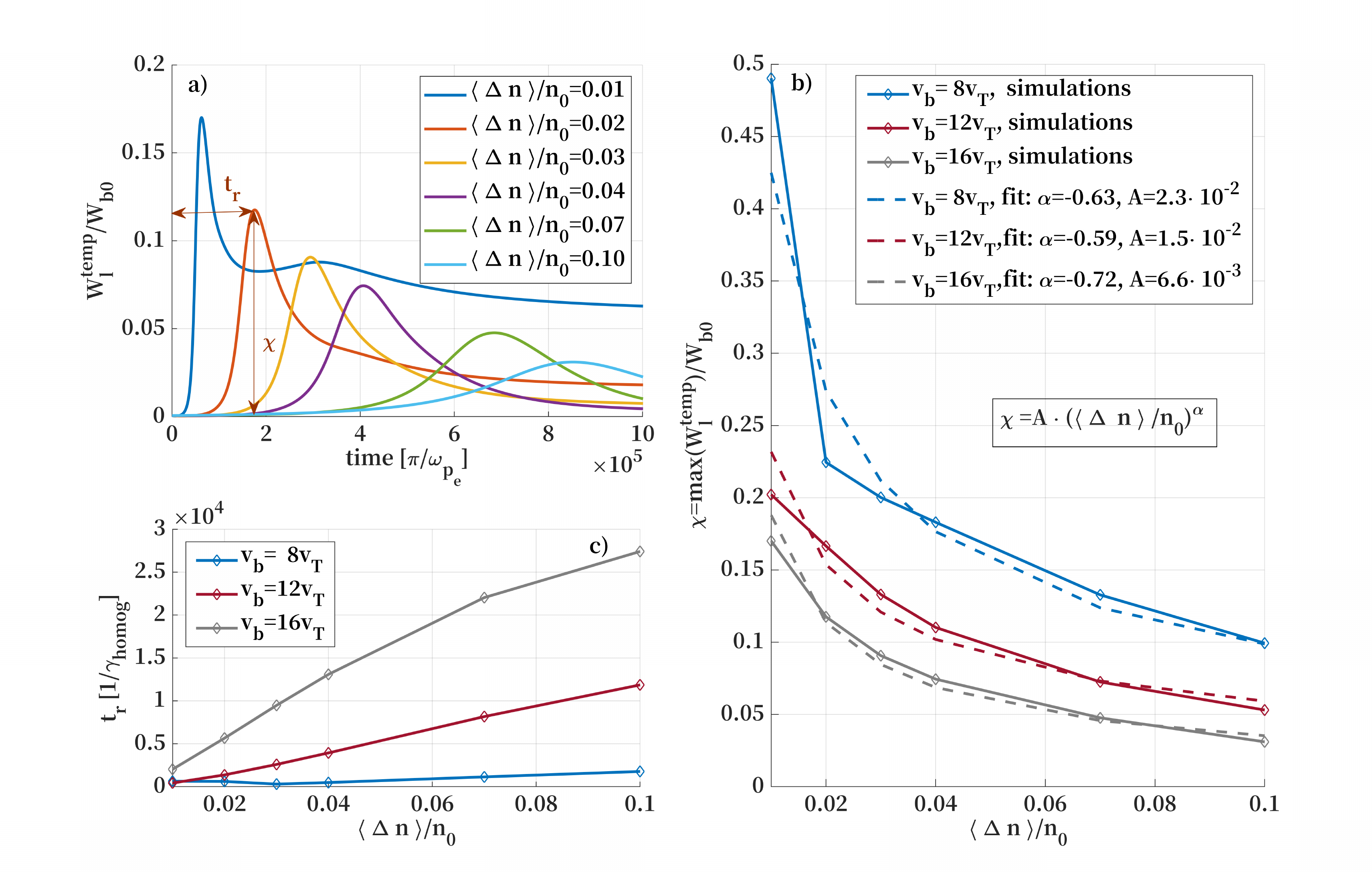}
\caption{Panel a) temporal evolution of the energy density of Langmuir waves $W^{temp}_l$ in units of initial energy density of electron beam $W_{b0}$, obtained from a probabilistic model \citep{voshchepynets2015probabilistic}. Numerical solutions
were found under the following conditions: electron density in the beam $n_{b}=10^{-5}n_{0}$, $v_{b}=16v_{T}$
and for six different levels of average density fluctuations ${\langle \Delta
n\rangle}/{n_{0}}$. Panel b) maximum values of Langmuir wave energy density (see panel a)) vs. average level of density fluctuations, for three different ratios of initial beam velocity $v_{b}$ and electron thermal velocity $v_{T}$. Simulation results (solid lines) are fitted by a power-law (dashed lines). Panel c) time of Langmuir waves energy growth during the beam relaxation (see panel a)) in units of growth rate in a homogeneous plasma $\gamma_{homog}$ vs. average level of density fluctuations, for three different ratios of initial beam velocity $v_{b}$ and electron thermal velocity $v_{T}$.}
\label{fig:fig_W_ev}
\end{figure}

It was shown by \cite{voshchepynets2015probabilistic2} that there exist two
regimes of beam-plasma interaction, depending on ratio of two important
parameters of the problem: the dispersion $k_{l}^{2}\lambda _{D}^{2}$ and the density fluctuation level ${\langle \Delta n \rangle}/{n_{0}}
$. When the level of the density fluctuations is small with respect to
dispersion effects, the relaxation occurs very similarly to the homogeneous case. In the opposite situation, the
dynamics of the instability is quite different. First of all, waves
grow much slower (Fig.\ref{fig:fig_W_ev}a,c) since the increment significantly decreases. Initially waves energy
grows, reaches a maximum and begins to decrease, as shown
on Fig.\ref{fig:fig_W_ev}a. The most surprising result is the transfer of a significant part of the wave energy to electrons with energies higher than
the energy of the beam.

Both descriptions -  probabilistic models and models based on Zakharov's equations, give very similar results \citep{voshchepynets2017statistics}. It was shown by \cite{voshchepynets2015probabilistic2} that the relaxation process consists of two stages. During the
first stage, the major relaxation process occurs and its characteristic time
may be determined as in conventional QLT. However, at the end of first stage the
system does not reach a stable state, only a marginally stable state, when the
increment of wave growth is 10$^{-4}-$10$^{-5}$ smaller than the initial increment. During the second stage, the system exists in this quasi-stable state, and still generates waves that are significantly above noise level, but
with much smaller amplitudes than during the first stage of relaxation. This allows to explain simultaneous observations of strong Langmuir waves and
positive slope on electron velocity distribution function at large
distances from the Sun \citep{lin1981energetic}. 

\cite{krasnoselskikh2019efficiency} recently showed that density
fluctuations may also change the mechanism of generation of the
radio emission at the fundamental frequency close to the local plasma frequency. Basically, the mechanism consists in a direct conversion of
electrostatic Langmuir waves to electromagnetic waves when the Langmuir wave is
reflected from density clump. The estimated efficiency of such a transformation  may become as large as $10^{-4}$ for density fluctuations of the order of several
percent.

The harmonic emission is produced via a nonlinear process and, consequently, its efficiency depends on
the Langmuir wave amplitude, which can be  evaluated with the help of a probabilistic model. Fig.\ref{fig:fig_W_ev} shows the evolution of the energy density of Langmuir wave,
generated in a randomly inhomogeneous plasma by an electron beam. Results are
provided for typical physical parameters in source region of solar type III radio
burst: beam electron density $n_{b}=10^{-5}n_{0}$, beam velocity $v_{b}=16v_{T}$; and for four levels of average density fluctuations: ${\langle \Delta n\rangle}/{n_{0}}= 0.01, 0.02, 0.03, 0.04, 0.07$ and $0.1$%
. Langmuir waves energy density is shown as a ratio of initial
energy density of the beam $W_{b0} = n_b m_e v_b^2 / 2$. It is worth mentioning
that the probabilistic model describes the temporal evolution of Langmuir waves,
while in the  present study the problem should be treated as a spatial one,
with a boundary condition corresponding to a continuous ejection of electron
beams. It is well known that the solutions of these two problems are rather
similar, except that the quasi-equilibrium saturation state corresponds to
a redistribution of energy fluxes rather than of the energy itself. This
implies a higher level of electrostatic waves energy in the spatial problem  with
respect to the temporal problem. The wave energy flux moves with the waves group
velocity, and the wave energy density may be found from the solution of the
temporal problem:
\begin{equation}
W^{SBP}_l=\frac{V_{ph}}{V_{gr}}W^{temp}_l, 
\end{equation}
where $W^{SBP}_l$ is the wave energy in the framework of the spatial boundary problem, while  $W^{temp}_l$ is the wave energy corresponding to
quasi-saturation in the framework of the temporal problem, $%
V_{gr}$ is the group velocity of Langmuir
waves. Since for the waves generated by the beam one has ${V_{ph}}/{%
V_{gr}}={v_{b}^{2}}/{3 v_{T}^{2}}$, it implies that for beams having velocities 
$10\div15 v_{T}$, ${V_{ph}}/{V_{gr}}$ may vary from 33 to 75. It also leads to an intensification of the waves in a relatively small region of space.

In order to estimate $E_0$, we hereafter use the maximum energy density of Langmuir waves $W_{l_{\max}} \equiv \max(W_{l})$, reached during the beam relaxation in a plasma with density fluctuations (see Fig.\ref{fig:fig_W_ev}a), obtained in the framework of the spatial boundary problem:
\begin{equation}
    W_{l_{\max}} = \max \left(W_{l_{\max}}^{SPB}\right) = \max \left( \frac{v_b^2}{3 v_{T}^2} W_{l_{\max}}^{temp} \right) = \frac{v_b^2}{3 v_{T}^2} \chi (\langle \Delta n \rangle / n_0) W_{b0},
\end{equation}
where $\chi$ is the characteristic coefficient that shows ratio of the maximal energy density reached during the relaxation process with respect to the initial energy density of the beam $W_{b0}$ (Fig.\ref{fig:fig_W_ev}b) at a given level of density fluctuations.

\section{Efficiency of conversion of Langmuir waves into harmonic EM emission} \label{sec:effic} %%%%%%%%%% Statistically

\begin{figure}[htbp]
\centering
\includegraphics[width = 0.48\textwidth]{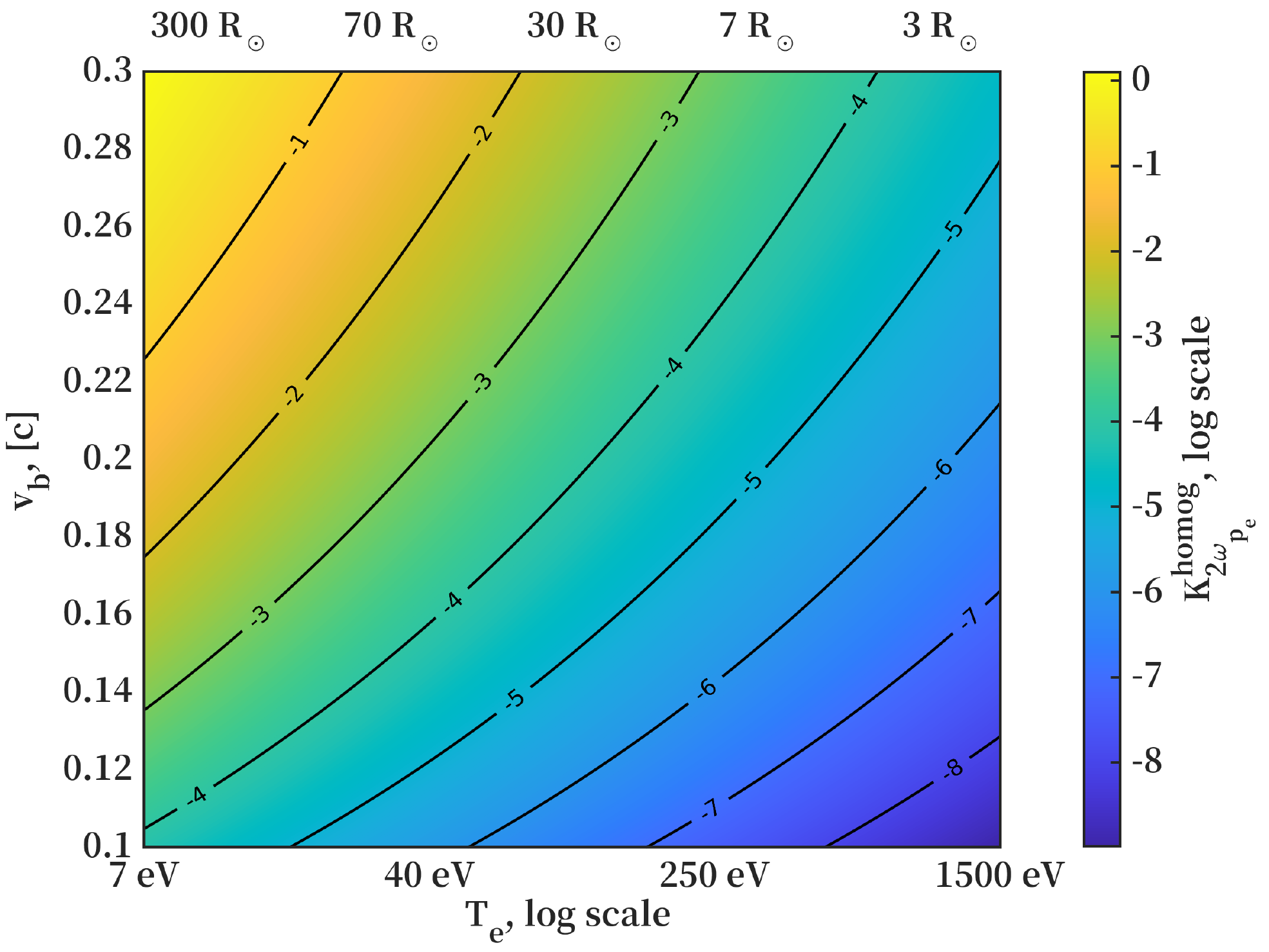} \includegraphics[width = 0.48\textwidth]{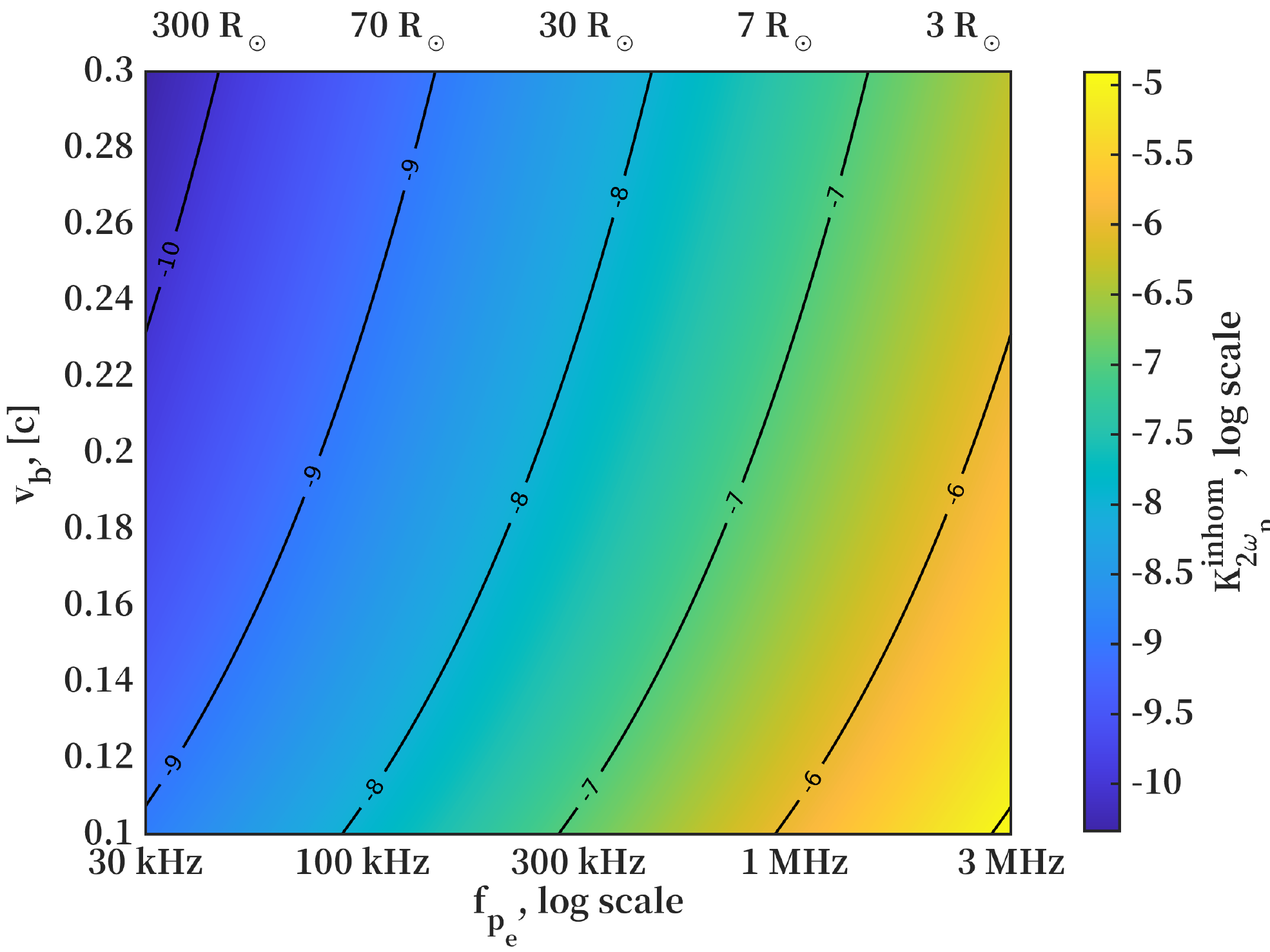} 
\caption{Efficiency of conversion of beam-generated Langmuir waves to harmonic EM emission in homogeneous (left panel) and inhomogeneous (right panel) plasma. The radial distance from the Sun in solar radii was inferred from temperature and ion density scaling for slow solar wind by \cite{meyer1998electron}. Left panel: $K_{2 \omega_{p_e}}^{homog}$ vs. electron beam velocity $v_b$ and electron temperature $T_e$. Parameters are: $\chi / \xi = 1$, $\Lambda = 10$, $P_{ref} = 0.5$, $\Delta v_b / v_b = 1/3$. Right panel: $K_{2 \omega_{p_e}}^{inhom}$ vs. electron beam velocity $v_b$ and electron plasma frequency $f_{p_e}$. Parameters are: $\epsilon = 0.75$, $P_{ref} = 0.5$, $n_b = 2.5 \cdot 10^{-5} n_0$, $T_e = 100$ eV, $\langle \Delta n \rangle / n_0 = 0.1$ (corresponding $\chi \approx 0.1$, $L_{sc} \approx 1.4\cdot 10^6$ cm).}
\label{fig: K_hom}
\end{figure}

Using the results of the previous section, one can
obtain $W_{l_{\max}}= ({v_b^2}/{v_{T}^2})\chi(\langle \Delta n \rangle/ n_0)n_{b}m_e v_{b}^{2} / 6$, and the linear increment of the growth of Langmuir waves may be estimated as $\gamma _{lin}=\xi (\langle \Delta n \rangle / n_0)\omega_{p_e}%
({n_{b}}/{n_0})({v_{b}^{2}}/{\Delta v_{b}^{2}})$, where $\chi (\langle \Delta n \rangle / n_0)
$ is the coefficient characterizing the ratio of the maximum energy density of Langmuir waves in an inhomogeneous plasma with respect to the initial electron beam energy and $\xi (\langle \Delta n \rangle / n_0)$ is the ratio of the increment of the instability in the inhomogeneous case with respect to the increment in the homogeneous case. In order to evaluate it in the computer simulations of beam-plasma interaction, we have made direct evaluation of the time of instability development, raising time $t_r$ for different beam velocities and levels of the density fluctuations as shown on the Fig.\ref{fig:fig_W_ev}. Values of the ${\chi(\langle \Delta n\rangle / n_0)} / {\xi(\langle \Delta n \rangle/ n_0)}$ typically vary from 10 to 70. The efficiency of conversion of Langmuir waves into harmonic emission in a quasihomogeneous plasma is%
\begin{equation}
K_{2 \omega_{p_e}}^{homog}=\frac{W_{t_{\max}}^{homog}}{W_{l_{\max}}}= \frac{\chi (\langle \Delta n \rangle / n_0)}{\xi
(\langle \Delta n \rangle / n_0)}\Lambda \frac{m_e \Delta v_{b}^{2}}{k_B T_e} \frac{P_{ref}(1-P_{ref})}{2400\pi ^{4}} \left(\frac{v_b}{v_{T}}\right)^2
\left( \frac{k_{b}}{\Delta k_b}\right) ^{4}\frac{\mathbf{k}_{t}^{5}}{k_{b}^{5}}.
\end{equation}
Taking into account that $k_{t}={\sqrt{3}\omega_{p_e}} / {2c}$, $k_b = {\omega_{p_e}} / {v_b}$ and ${\Delta k_b} / {k_{b}} \simeq {\Delta v_{b}} / {v_{b}}$, we obtain 
\begin{equation}
K_{2 \omega_{p_e}}^{homog}=\frac{\sqrt{3}}{38400 \pi^4} \Lambda P_{ref}(1-P_{ref}) \frac{\chi (\langle \Delta n \rangle/ n_0)}{\xi
(\langle \Delta n \rangle/ n_0)} \left(\frac{v_{b}}{c}\right)^9 \left( \frac{v_{b}}{\Delta v_{b}}\right) ^{2} \left(\frac{m_e c^2}{k_B T_e}%
\right)^{2}.
\end{equation}%
As a next step, we estimate the efficiency of conversion of a Langmuir wave into harmonic EM emission in the vicinity of the reflection points: 
\begin{eqnarray}
K_{ \ 2 \omega_{p_e}}^{inhom} = \frac{\langle W_t^{inhom}\rangle_{\psi, \delta n, L} }{W_{l_{\max}}} = \frac{25 \sqrt{6}}{3} \epsilon \chi (\langle \Delta n \rangle / n_0) P_{ref} \frac{n_b}{n_0} \frac{v_b}{c} \left(\frac{v_{T}}{c} \right)^2 \left({\frac{k_l^2 \lambda_D^2}{\langle \Delta n \rangle / n_0}}\right)^{2} \frac{\omega_{p_e}^2 L_{sc}^2}{c^2}.
\label{eq:K_harm}
\end{eqnarray}
The dependence of the efficiency coefficients $K^{homog}_{2\omega_{p_e}}$ and $K^{inhom}_{2\omega_{p_e}}$ on the plasma temperature and electron plasma frequency respectively vs. electron beam velocity is presented in Figure \ref{fig: K_hom}.

\section{Radiation Directivity and Intensity in the Solar Corona and Wind} \label{sec:discuss} %%%%%%%%%%%%%%%%%%Discuss

\subsection{Radiation Directivity}

The directivity of harmonic radiation of type III radio bursts in a homogeneous plasma has been extensively discussed shortly after the suggestion of the plasma emission mechanism by \cite{ginzburg1958possible}. According to this mechanism, the harmonic emission is always quadrupolar, its angular range of visibility depends on the value of beam velocity, and is typically larger for smaller values of beam velocity \citep{zheleznyakov1970theory}. Disregarding the revisions of the initial theory, the general directivity characteristics, predicted for such emission remain unchanged.   

\begin{figure}[htbp]
\centering
\includegraphics[width = 0.4\textwidth]{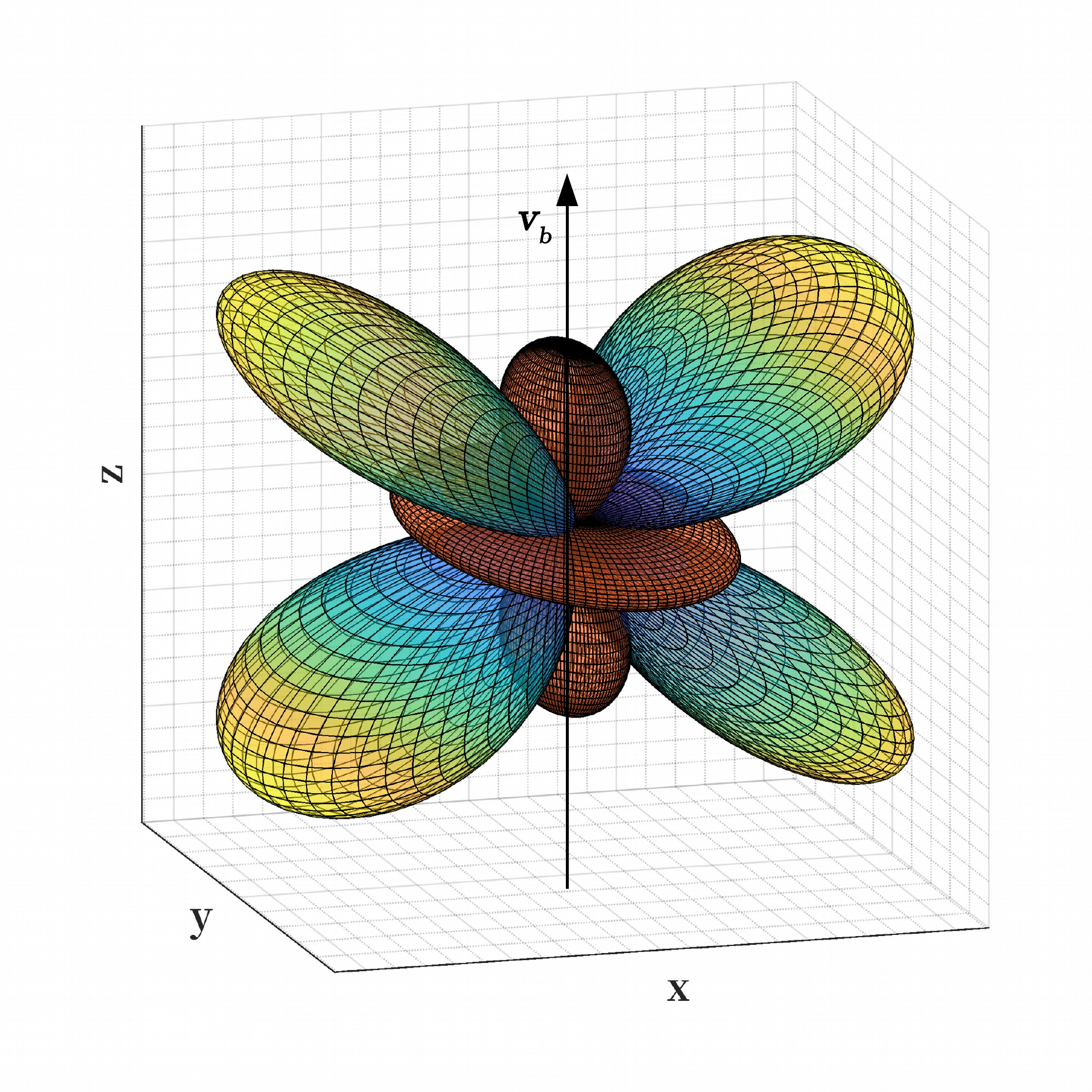}
\caption{A schematic radiation pattern of harmonic EM emission, produced via the plasma emission mechanism in homogeneous plasma (multi-colored) and via Langmuir wave coalescence inside density clumps (orange). Here the ratio between homogeneous and inhomogeneous parts is chosen to be 5:1.}
\label{fig:rad}
\end{figure}

 In the present paper, we have revisited the emission mechanism in a quasihomogeneous plasma and also have considered a physical process of emission from localized regions (clumps), where the reflection of Langmuir waves occurs. Similarly to the plasma emission mechanism, this harmonic radiation is also quadrupolar. But, unlike the plasma emission mechanism, it produces EM emission in the parallel and perpendicular directions with respect to electron beam direction (see Fig.\ref{fig:rad}). In this context, we can deduce that the direction of the density gradient inside the density clump and the direction of electron beam should be roughly similar based on the following conditions: (1) the angular range of directivity of Langmuir waves, generated by electron beam, is quite narrow: up to $20^{o}$ with respect to beam direction \citep{krasnoselskikh2011determining}, (2) the harmonic emission in our model can be generated only for an incident Langmuir wave that is highly aligned with density gradient direction, since the angle of incidence $\psi$, that allows successful production of harmonic, is strictly determined by beam velocity: $\psi \approx {\sqrt{3}} v_b / 2 c $, and is small ($\lesssim 10^{o}$). 
 
We can schematically compare the two aforementioned radiation patterns by positioning them against the direction of electron beam (Fig.\ref{fig:rad}). Here we do not precisely represent the magnitudes of relevant intensity or angular range of radiation, but rather to give a formal comparison of the major emission directions, given by these two mechanisms combined. Disregarding that propagation effects, such as refraction, absorption and reflection (of backward-emitted emission) cause a widening of the angular range, where the harmonic emission from  homogeneous plasma is visible, they are insufficient to explain a widespread visibility of harmonic emission of type III radio bursts \citep{thejappa2007monte}. As such, the mechanism of generation of radio emission inside density clumps adds parallel and perpendicular-directed radiation to the conventional harmonic radiation pattern and contributes to the general visibility of harmonic emission of type IIIs. 

\subsection{Radiation Intensity in the Corona and Solar Wind}

Observations of type III radio bursts indicate that fundamental-harmonic pairs represent the majority of radio bursts in the high frequency range. The fundamental usually begins below 100 MHz, while the harmonic can begin as high as $\sim$ 500 MHz \citep{dulk1980position}. At the same time, the rarity of the fundamental emission in the 100-500 MHz range remains unexplained, as the absorption due to inverse bremsstrahlung becomes significant only above around 500 MHz for the fundamental and above 1 GHz for the harmonic emission \citep{reid2014review}. This might indicate that in the corona and its proximity the efficiency of  conversion of Langmuir waves into electromagnetic radio emission is higher for certain mechanisms of harmonic radiation generation. Further, we will formally discuss intensity of the radiation, remembering its simple relation to the wave energy density $I \sim W_t$. 

We have revisited a well-known result for harmonic emission from a quasihomogeneous plasma and obtained an analytical result that is different from the one obtained by \cite{willes1996second}. We showed that the assumptions used by \cite{willes1996second} are not always justified and they lead to a sufficient underestimation of the EM wave amplitude. We have performed here a direct calculations for the general case. According to our results, the intensity of such emissions is much higher than previously predicted. Such emission is more efficiently produced for a larger ratio of $v_b$ to $c$ and smaller electron temperatures $T_e$ (see the left panel of  Fig.\ref{fig: K_hom}). As electron temperature is decreasing with heliocentric distance, we can infer a dependence of $T_e$ on radial distance from the Sun, making use of one of solar wind models (e.g., \cite{meyer1998electron}), and make a comparison between the parameter domains of domination of harmonic emissions from a homogeneous plasma and from density clumps.

The harmonic emission from density clumps is the most intense at smaller heliocentric distances. Its efficiency of generation is higher for smaller ratios of $v_b$ to $c$ and for a smaller level of density fluctuations (see the right panel of Fig.\ref{fig: K_hom}). Here we note that we have applied a constant level of electron temperature $T_e$ and density fluctuations throughout the whole plasma frequency interval. Recent studies show that, according to \textit{in situ} measurements of Parker Solar Probe, the level of density fluctuations at around 36 $R_\odot$ is about 0.06-0.07, and it is predicted to grow up to $\sim 0.2$ at a distance of a few solar radii \citep{krupar2020density}. On the right panel of Fig.\ref{fig: K_hom} we implied a level of density fluctuations $\langle \Delta n \rangle / n_0 = 0.1$.  

As we compare the efficiency of conversion of Langmuir waves into harmonic emission and thus indirectly the relevant intensities, we see that emission from density clumps can become as important as the emission from quasihomogeneous plasma at around a few solar radii, closer to the low corona. 

\subsection{Summary}

It is widely accepted that electron density fluctuations in the solar wind affect the propagation of radio emission. On the other hand, there are very few studies of the impact of these inhomogeneities on the process of generation of such radio emissions. In present paper, we have considered the generation of harmonic radio emission via $l + l' \rightarrow t$ process under two different circumstances: in a quasihomogeneous plasma (through coalescence of two nearly oppositely propagating Langmuir waves) and inside structures, formed by density fluctuations with increasing density gradient (via coalescence of a Langmuir wave with its reflected part in the vicinity of the reflection point). For the first generation process we have made the following assumptions: (1) coalescence takes place in a homogeneous plasma, (2) the spectrum of forward moving and reflected Langmuir waves is Gaussian, (3) the population of reflected waves is the result of the reflection of a part of forward moving waves from density irregularities, the reflection process is taken into account by means of a coefficient $P_{ref}$, (4) two coalescing Langmuir waves meet head-on. For the second generation process, we used the system of equations proposed by \cite{zakharov1972collapse}. The assumptions we made when deriving the solution are: (1) $3 k_l^2 \lambda_D^2 \ll \langle \Delta n \rangle / n_0$, i.e., linear dispersion is less significant compared to the effect of density fluctuations, (2) $k_l L_{sc}\gg 1$, i.e., characteristic scales of density gradients inside density clumps are significantly larger than the wavelength of the Langmuir wave, (3) $\psi \ll 1$, i.e., the incident Langmuir wave should be closely aligned with the direction of the density gradient and, consequently, almost anti-parallel to the reflected wave (HOA), (4) clumps are approximately spherical with the electron density increasing linearly towards their center, (5) the radio emission is formed within the conversion region, (6) the quadrupole component of the harmonic emission is dominant. In both cases, the plasma is unmagnetized and  we consider only the conversion of Langmuir waves into \textit{harmonic} emission. This way we have obtained analytical expressions for the energy density of harmonic radio emission in both cases. And finally, we have estimated the efficiency of conversion of beam-generated Langmuir waves into harmonic electromagnetic emission from both regions of emission.

\section{CONCLUSIONS}

(1) a direct calculation of the generation of harmonic EM emission via the process of coupling of primary beam-generated Langmuir wave with the reflected wave in a quasihomogeneous plasma yields a higher radiation intensity than found  previously (e.g., by \cite{willes1996second}),

(2) a new model of generation of harmonic emission inside density clumps close to the region of reflection of Langmuir waves demonstrates the efficiency of conversion of Langmuir waves into EM waves, which is under certain conditions comparable with the aforementioned quasihomogeneous plasma emission and even prevails at smaller heliocentric distances,

(3) EM radiation from density clumps may be important for the visibility of harmonic emissions.

\acknowledgements{VK acknowledges the financial support from CNES through grants "Search Coil for Solar Orbiter" and "Parker Solar Probe" and financial support from NASA through the grant 80NSSC20K0697. Authors are grateful to Didier Mourenas for useful discussions.} 

\appendix
\section{Nonlinear coupling of Langmuir waves in a quasihomogeneous plasma} \label{app: 1}

The process of EM waves generation by the coupling of Langmuir waves is
described by the following set of equations \citep{tsytovich2012nonlinear}:

\begin{equation}
\frac{dN_{t}(\mathbf{k}_{t})}{dt}=\int \int \frac{d^{3}\mathbf{k}_{1}d^{3}%
\mathbf{k}_{2}}{(2\pi )^{6}}w_{ll}^{t}(\mathbf{k}_{1},\mathbf{k}_{2},\mathbf{k}_{t})[N_{l}(%
\mathbf{k}_{1})N_{l}(\mathbf{k}_{2})-N_{l}(\mathbf{k}_{1})N_{t}(\mathbf{k}%
_{t})-N_{l}(\mathbf{k}_{2})N_{t}(\mathbf{k}_{t})], 
\label{eq: number_of_quanta}
\end{equation}

\begin{equation}
w_{ll}^{t}(\mathbf{k}_{1},\mathbf{k}_{2},\mathbf{k}_{t})=\frac{(2\pi
)^{6}e^{2}(k_{1}^{2}-k_{2}^{2})^{2}}{32\pi m_e^{2}\omega_{p_e}k_{t}^{2}}\frac{[%
\mathbf{k}_{1}\times \mathbf{k}_{2}]^{2}}{k_{1}^{2}k_{2}^{2}}\delta (\omega
_{t}-\omega _{1}-\omega _{2})\delta (\mathbf{k}_{t}-\mathbf{k}_{1}-\mathbf{k}%
_{2}). 
\label{eq: matrix_element}
\end{equation}

For our problem the number of quanta of Langmuir waves is supposed to be much larger than the number of quanta of EM waves, thus the Eq.(\ref{eq: number_of_quanta}) may be simplified as follows:

\begin{equation}
\frac{dN_{t}(\mathbf{k}_{t})}{dt}=\int \int \frac{d^{3}\mathbf{k}_{1}d^{3}%
\mathbf{k}_{2}}{(2\pi )^{6}}w_{ll}^{t}(\mathbf{k}_{1},\mathbf{k}_{2},\mathbf{k}_{t})N_{l}(%
\mathbf{k}_{1})N_{l}(\mathbf{k}_{2}). 
\label{eq: number_of_quanta_simplified}
\end{equation}

Aiming to evaluate the number of quanta of EM waves we choose the distribution 
to consist of two symmetric parts, primarily generated Langmuir waves and
reflected that have similar distribution functions, but different intensities. Similarly to \cite{willes1996second}, we choose both distributions to be Gaussian and centered at $\mathbf{k}_b$ ($k_b = \omega_{p_e} / v_b$) for forward moving waves and at $-\mathbf{k}_b$ for reflected waves. This suggests that reflections are of the mirror type. Under such conditions the total distribution can be written as follows:
\begin{equation}
N_l \left( k_{\parallel },k_{\perp }\right) =(1-P_{ref})N \exp (-\frac{%
(k_{\parallel }-k_{b})^{2}+k_{\perp }^{2}}{\Delta k^{2}})+P_{ref} N \exp (-\frac{%
(k_{\parallel }+k_{b})^{2}+k_{\perp }^{2}}{\Delta k^{2}}), 
\end{equation}
where N is normalized accordingly
\begin{equation}
N =\frac{1}{\pi ^{3/2}\Delta k^{3}} \frac{W_l}{\omega_{p_e}}.
\end{equation}
This way the Eq.(\ref{eq: number_of_quanta_simplified}) takes the form:
\begin{equation}
\begin{split}
\frac{dN_{t}(\mathbf{k}_{t})}{dt}=\frac{e^{2}P_{ref}(1-P_{ref})N^{2}}{32\pi m_e^{2}\omega
_{pe}(\pi ^{3}\Delta k^{6})}\times 
\\
\iint d^{3}\mathbf{k}_{1}d^{3}\mathbf{k}_{2}\frac{%
(k_{1}^{2}-k_{2}^{2})^{2}}{k_{t}^{2}}\frac{[\mathbf{k}_{1}\times \mathbf{k}%
_{2}]^{2}}{k_{1}^{2}k_{2}^{2}}\delta (\omega _{t}-\omega _{1}-\omega
_{2})\delta (\mathbf{k}_{t}-\mathbf{k}_{1}-\mathbf{k}_{2})\exp \left[-\frac{%
(k_{1\parallel }-k_{b})^{2}+k_{1\perp }^{2}}{\Delta k^{2}}\right] \exp \left[-\frac{%
(k_{2\parallel }+k_{b})^{2}+k_{2\perp }^{2}}{\Delta k^{2}}\right]. 
\end{split}
\label{eq: number_of_quanta_with_spectra }
\end{equation} 

\begin{figure}[htbp]
\centering
\includegraphics[width = 0.3\textwidth]{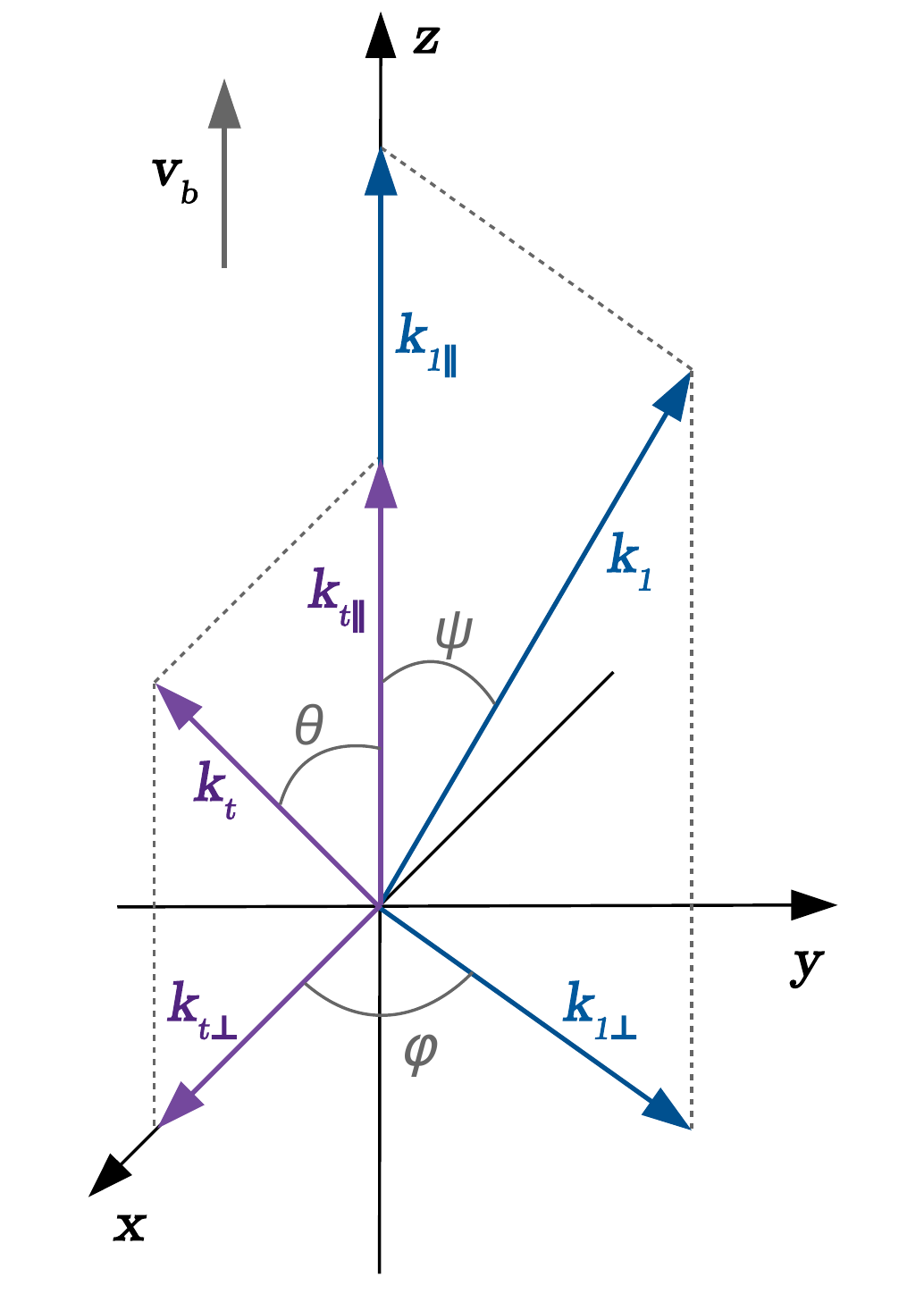}
\caption{Schematic illustration of wavevectors $\mathbf{k}_1$ and $\mathbf{k}_t$ against the direction of the electron beam ($z$-axis).}
\label{fig:k_vectors_illustration}
\end{figure}

We use the reference frame where the $z$-axis is in a parallel direction along the direction of propagation of the beam that is generating primary Langmuir waves (or along the magnetic
field that is not taken into account here but still present in the solar
wind). The $x$-axis is a perpendicular direction that is chosen to be along the second component of the $\mathbf{k_{t}}$ vector. Three vectors $\mathbf{k}_{t},\mathbf{k}_{1}$ and $\mathbf{k}_{2}$
make the triangle (see Fig.\ref{fig:k_vectors_illustration}) according to momentum conservation so that
\begin{equation}
\mathbf{k}_{2}=\mathbf{k}_{t}-\mathbf{k}_{1}.
\end{equation}

The natural assumption here is $|\mathbf{k}_{t}| \ll |\mathbf{k}_{1, 2}|$, thus 
\begin{equation}
\mathbf{k}_{2}^{2}=\mathbf{k}_{1}^{2}-2k_{t\parallel }k_{1\parallel
}-2k_{1\perp }k_{t\perp }\cos \varphi \simeq \mathbf{k}_{1}^{2}-2k_{t}k_{1}(\cos
\psi \cos \theta +\sin \psi \sin \theta \cos \varphi ),
\end{equation}
here $\psi$ and $\theta$ are the angles between $z$-axis and vectors $\mathbf{k}_{1}$ and $\mathbf{k}_{t}$ respectively, $\varphi $ is the angle between the projections of $\mathbf{k}_{1}$ and $\mathbf{k}_{t}$ to the plane perpendicular to $z$-axis ($xy$-plane).
The vector $\mathbf{k}_{t}$ without loss of generality is chosen to be in
the $xz$ plane. Consequently the multiplier from Eq.(\ref{eq: number_of_quanta_with_spectra }) may be rewritten as

\begin{equation}
\frac{(k_{1}^{2}-k_{2}^{2})^{2}}{k_{t}^{2}}\frac{[\mathbf{k}_{1}\times 
\mathbf{k}_{2}]^{2}}{k_{1}^{2}k_{2}^{2}}= 4 k_t^2 \times R(\psi ,\theta ,\varphi ),
\end{equation}
where  the trigonometric expression $R(\psi ,\theta ,\varphi )$ stands for
\begin{equation}
\begin{split}
R(\psi ,\theta ,\varphi )= \\
=(\sin ^{2}\psi \sin ^{2}\theta \cos ^{2}\varphi +2\sin \psi \cos \psi \sin
\theta \cos \theta \cos \varphi +\cos ^{2}\psi \cos ^{2}\theta )\times \\
\lbrack -\sin ^{2}\theta \sin ^{2}\psi \cos ^{2}\varphi -2\sin \theta \cos
\theta \sin \psi \cos \psi \cos \varphi +(\cos ^{2}\psi \sin ^{2}\theta +\sin
^{2}\psi )].
\end{split}
\end{equation}

After the integration of Eq.(\ref{eq: number_of_quanta_with_spectra }) over $\mathbf{k}_{2}$ we obtain
\begin{equation}
\frac{dN_{t}(\mathbf{k}_{t})}{dt}=\frac{e^{2}P_{ref}(1-P_{ref})N^{2} \mathbf{k}_{t}^2}{8\pi m_e^{2}\omega
_{pe}(\pi ^{3}\Delta k^{6})} \times G,
\end{equation}
here we used a denotation
\begin{eqnarray}
G= \int\limits_{-\pi }^{\pi }d\varphi \int\limits_{0}^{\pi
}\sin \psi d\psi \int k^{2}dk\delta (\omega _{t}-\omega _{1}-\omega
_{2})R(\psi ,\theta ,\varphi ) \times \nonumber\\
\exp (-\frac{(k_{1\parallel
}-k_{b})^{2}+k_{1\perp }^{2}}{\Delta k^{2}})\exp (-\frac{(k_{t\parallel
}-k_{1\parallel }+k_{b})^{2}+(k_{1\perp }^{2}+k_{t\perp }^{2}-2k_{1\perp
}k_{t\perp }\cos \varphi )}{\Delta k^{2}}). 
\end{eqnarray}
The first integration over $\mathbf{k}_{1}$ should take into account the delta function
over frequencies. In the first order
approximation, neglecting terms that are linear on $\mathbf{k}_{t}$, the argument of the delta function may be rewritten as follows 
\begin{equation}
\omega _{t}-\omega _{1}-\omega _{2}=\omega _{t}-2\omega_{p_e}-3\omega
_{p_e}\lambda _{D}^{2}k_{1}^{2}=-3\omega_{p_e}\lambda
_{D}^{2}(k_{1}^{2}-k_{r}^{2}), 
\end{equation}%
where 
\begin{equation}
k_{r}^{2}=\frac{\omega _{t}-2\omega_{p_e}}{3\omega_{p_e}\lambda _{D}^{2}}. 
\end{equation}%
Thus the delta function has the form 
\begin{equation}
\delta (\omega _{t}-\omega _{1}-\omega _{2})=\frac{1}{6\omega_{p_e}\lambda
_{D}^{2}k_{r}}\delta (k_{1}-k_{r}), 
\end{equation}%
and consequently 
\begin{equation}
G=\frac{k_{r}}{6\omega_{p_e}\lambda _{D}^{2}}\int\limits_{-\pi }^{\pi }d\varphi
\int\limits_{0}^{\pi }\sin \psi d\psi R(\psi ,\theta ,\varphi )\exp (-%
\frac{2(k_{r}^{2}-2k_{r}k_{b}\cos \psi +k_{b}^{2})}{\Delta k^{2}}).
\end{equation}

The expression $R(\psi ,\theta ,\varphi )$ may be
rewritten as follows:%
\begin{equation}
\begin{split}
R\left( \theta ,\psi ,\varphi \right) =-\frac{1}{8}\cos 4\varphi \sin ^{4}\theta
\sin ^{4}\psi - 
\\
-\cos 3\varphi \sin ^{3}\theta \cos \theta \sin ^{3}\psi \cos \psi + 
\\
+\frac{1}{2}\cos 2\varphi \lbrack -\sin ^{4}\theta \sin ^{4}\psi +\sin ^{4}\theta
\sin ^{2}\psi \cos ^{2}\psi +\sin ^{4}\psi \sin ^{2}\theta -5\sin
^{2}\theta \cos ^{2}\theta \sin ^{2}\psi \cos ^{2}\psi ]+ 
\\
+\cos \varphi \lbrack -3\sin ^{3}\theta \cos \theta \sin ^{3}\psi \cos \psi
-2\sin \theta \cos ^{3}\theta \sin \psi \cos ^{3}\psi +2\sin ^{3}\theta \cos
\theta \sin \psi \cos ^{3}\psi +2\sin \theta \cos \theta \sin ^{3}\psi \cos
\psi ]+ 
\\
-\frac{3}{8}\sin ^{4}\theta \sin ^{4}\psi +\frac{1}{2}[\sin ^{4}\theta \sin
^{2}\psi \cos ^{2}\psi +\sin ^{4}\psi \sin ^{2}\theta -5\sin ^{2}\theta
\cos ^{2}\theta \sin ^{2}\psi \cos ^{2}\psi ]+(\cos ^{2}\psi \sin
^{2}\theta +\sin ^{2}\psi )\cos ^{2}\psi \cos ^{2}\theta. 
\end{split}
\end{equation}

Next step consists in integration over $d\varphi $, taking into account the following relation
\begin{equation}
\int\limits_{0}^{2\pi }d\varphi \cos n\varphi \exp (Z\cos \varphi )=I_{n}(Z), 
\end{equation}
here $I_n(Z)$ is a modified Bessel function of the $n$-th order.We should point out a very important difference of our calculation with the
one by \cite{willes1996second}. The assumption made by \cite{willes1996second} consists in
inequality 
\begin{equation}
\frac{2k_{r}k_{t}\sin \psi \sin \theta }{\Delta k^{2}}\gg1, 
\end{equation}%
and as will be seen later it is not satisfied for this calculation. Indeed, the inequality 
\begin{equation}
    \frac{2 k_r k_t}{\Delta k^2} \gg 1
\end{equation}
may be satisfied since $\Delta k \ll k_r$ (the spectrum of generated waves may be considered to be rather narrow), but the multiplier $\sin \psi \sin \theta$ may be quite small as we use a head-on approximation that naturally comes from our problem statement. Integration over $d\varphi$ results in
\begin{equation}
\begin{split}
G=\frac{k_{r}}{6\omega_{p_e}\lambda _{D}^{2}}%
 \int\limits_{0}^{\pi }\sin \psi d\psi \exp (-\frac{%
2(k_{r}^{2}-2k_{r}k_{b}\cos \psi +k_{b}^{2})}{\Delta k^{2}})\exp (-\frac{%
2k_{t}\cos \theta (k_{b}-k_{r}\cos \psi )}{\Delta k^{2}}) \times
\\
\{-\frac{1}{8}\sin ^{4}\theta \sin ^{4}\psi   I_{4}(\Xi)
-\sin ^{3}\theta \cos \theta \sin ^{3}\psi \cos \psi   I_{3}(\Xi)+
\\
+\frac{1}{2}[-\sin ^{4}\theta \sin ^{4}\psi +\sin ^{4}\theta \sin ^{2}\psi
\cos ^{2}\psi +\sin ^{4}\psi \sin ^{2}\theta -5\sin ^{2}\theta \cos ^{2}\theta
\sin ^{2}\psi \cos ^{2}\psi ]  I_{2}(\Xi)+
\\
+[-3\sin ^{3}\theta \cos \theta \sin ^{3}\psi \cos \psi -2\sin \theta \cos
^{3}\theta \sin \psi \cos ^{3}\psi +2\sin ^{3}\theta \cos \theta \sin \psi
\cos ^{3}\psi +2\sin \theta \cos \theta \sin ^{3}\psi \cos \psi ] 
I_{1}(\Xi)+
\\
\lbrack -\frac{3}{8}\sin ^{4}\theta \sin ^{4}\psi +\frac{1}{2}[\sin ^{4}\theta
\sin ^{2}\psi \cos ^{2}\psi +\sin ^{4}\psi \sin ^{2}\theta -5\sin
^{2}\theta \cos ^{2}\theta \sin ^{2}\psi \cos ^{2}\psi ]+(\cos ^{2}\psi
\sin ^{2}\theta +\sin ^{2}\psi )\cos ^{2}\psi \cos ^{2}\theta ]  I_{0}(\Xi) \},
\end{split}
\label{eq: G}
\end{equation}
where $\Xi = \frac{2k_{r}k_{t}\sin \psi \sin \theta }{\Delta k^{2}}$. The last step in our calculation consists in calculation of the following
integral
\begin{equation}
\int\limits_{-1}^{1}d(\cos \psi )\exp (\frac{4 k_{r}k_{b}\cos \psi }{%
\Delta k^{2}})g(\cos \psi ),
\end{equation}%
where the factor ${4 k_{r}k_{b}} / {\Delta k^{2}}$ is very large. The standard asymptotic estimation of such integral, when the function has the maximum on the given interval and when the parameter $Y \gg 1$ is, according to \cite{wasow2018asymptotic} 
\begin{equation}
\int\limits_{-1}^{1}dt \exp (Yt) g(t)dt=\frac{1}{Y}\exp(Y) g(1) + O(Y^{-2}). 
\end{equation}%
Accordingly the Eq.(\ref{eq: G}) reduces to
\begin{equation}
G=\frac{\Delta k^{2}}{6 \omega_{p_e}\lambda _{D}^{2}k_{b}}\sin
^{2}\theta \cos ^{2}\theta \exp [-\frac{2(k_{r}-k_{b})^{2}}{\Delta k^{2}}-\frac{%
2k_{t}\cos \theta (k_{b}-k_{r})}{\Delta k^{2}}].
\end{equation}%
Taking into account that the maximum of this expression corresponds to $k_r = k_b$, we will estimate the result as
\begin{equation}
\frac{dN_{t}(\mathbf{k}_{t})}{dt}=\frac{P_{ref}(1-P_{ref})\omega_{p_e}^{2}}{{768}\pi ^{5}}%
\left( \frac{k_{b}}{\Delta k}\right) ^{4}\frac{\mathbf{k}_{t}^{2}}{k_{b}^{5}}\frac{%
N_{l}^{2}}{n_0 m_e v_{T}^{2}}\sin ^{2}\theta \cos ^{2}\theta,
\end{equation}%
or
\begin{equation}
\frac{dW_{t}^{homog}(\mathbf{k}_{t})}{dt}=\frac{P_{ref}(1-P_{ref})\omega_{p_e}}{{768}\pi ^{5}}%
\left( \frac{k_{b}}{\Delta k}\right) ^{4}\frac{\mathbf{k}_{t}^{2}}{k_{b}^{5}} \frac{%
W_{l}}{n_0 k_B T_e}  W_{l} \sin ^{2}\theta \cos ^{2}\theta.
\end{equation}%
In order to evaluate the characteristic energy of
electromagnetic wave, we integrate over $d^{3}\mathbf{k}_{t}$ 
\begin{equation}
\frac{dW_{t}^{homog}}{dt}=\frac{P_{ref}(1-P_{ref})\omega_{p_e}}{{7200 \pi ^{4}}}\left( \frac{%
k_{b}}{\Delta k}\right) ^{4}\frac{\mathbf{k}_{t}^{5}}{k_{b}^{5}} \frac{W_{l}}{n_0 k_B T_e}%
 W_l.
\end{equation}%
As the temporal evolution of Langmuir wave energy density may be presented in following form
\begin{equation}
W_l(t)=W_{noise}\exp (\gamma t),
\end{equation}%
total energy density of EM harmonic emission will be 
\begin{equation}
W_{t}^{homog}=\frac{P_{ref} (1-P_{ref})\omega_{p_e}}{{7200 \pi^4}\gamma }\left( \frac{%
k_{b}}{\Delta k}\right) ^{4}\frac{\mathbf{k}_{t}^{5}}{k_{b}^{5}}\frac{W_l}{%
n_0 k_B T_e} W_l.
\end{equation}%
A conventional estimate of the time of growth, namely $\frac{1}{\gamma }$
will be 
\begin{equation}
\frac{1}{\gamma }=\frac{\Lambda }{_{\gamma _{lin}}},
\end{equation}%
here $\gamma _{lin}$ is the linear increment of the instability of Langmuir
waves. 

\section{Equations for electrostatic potential, electric field and current density}   \label{app: 2}%%%%%%%%%%%%%%%%%%App 2

We start with the equation system for plasma oscillations with $k_l \lambda_D \ll 1$, derived by \cite{zakharov1972collapse}:

\begin{eqnarray} \label{eq:pyth}
\Delta \Phi_{l}^{inhom} (\bm{r},t) = 4 \pi e \delta n_l, \\
\frac{\partial }{\partial t} \delta n_l + \nabla ((n_0 + \delta n) 
\bm{v_e}) = 0, \\
\frac{\partial \bm{v_e}}{\partial t} = \frac{e}{m_e} \nabla \Phi_{l}^{inhom} (\bm{r},t)
- 3 v_{T}^2 \nabla \frac{\delta n_l}{n_0},
\end{eqnarray}
where electron density has the following form:
 
\begin{displaymath}
n_e = n_0 + \delta n + \delta n_l.
\end{displaymath}

Applying a method of small perturbations to our system and performing simple transformations, we obtain

\begin{eqnarray}
\frac{\partial^2}{\partial t^2} \frac{\Delta \Phi_{l}^{inhom} (\bm{r},t)}{4 \pi e} + n_0 \cdot \left(\frac{e%
}{m } \Delta \Phi_{l}^{inhom} (\bm{r},t) - \frac{3 v_{T}^2}{4 \pi e n_0 } \Delta
(\Delta \Phi_{l}^{inhom} (\bm{r},t)) \right) + \nonumber \\
+ \nabla \left(\delta n \left(\frac{e}{m_e} \nabla \Phi_{l}^{inhom} (\bm{r},t)
- \frac{3 v_{T}^2}{4 \pi e n_0} \nabla (\Delta \Phi_{l}^{inhom} (\bm{r},t)) \right)\right)= 0.
\label{eq:eq_for_Phi}
\end{eqnarray}

Bearing in mind that each quantity in Eq.(\ref{eq:eq_for_Phi}), except $\Phi_{l}^{inhom} (\bm{r},t)$ is assumed to be independent on spatial coordinates or to be a very slowly varying function of those comparatively to $\Phi_{l}^{inhom} (\bm{r},t)$, we rewrite the Eq.(\ref{eq:eq_for_Phi}): 

\begin{displaymath}
\Delta \left(-\frac{\omega_l^2}{4 \pi e } \Phi_{l}^{inhom} (\bm{r},t) + \frac{e n_0}{m} \Phi_{l}^{inhom} (\bm{r},t) - \frac{3 v_{T}^2}{4 \pi e} \Delta \Phi_{l}^{inhom} (\bm{r},t) + \frac{e \delta n}{m} \Phi_{l}^{inhom} (\bm{r},t) - \frac{3 v_{T}^2\delta n}{4 \pi e n_0} \Delta \Phi_{l}^{inhom} (\bm{r},t) \right) = 0.
\end{displaymath}
or, after a few transformations, 

\begin{equation}
\left(\omega_l^2 - \omega_{p_e}^2 \left(1+\frac{\delta n}{n_0} \right) + 3 v_{T}^2 \Delta \right) \Phi_{l}^{inhom} (\bm{r},t) = 0.
\end{equation}

The density gradient profile within the clump is assumed to be linear, directed along $z$-axis, and may be expressed by means of Heaviside step function $\theta (z)$:

\begin{displaymath}
\frac{\delta n}{n_0} = \theta (z) \frac{z}{L}.
\end{displaymath}

Now, since there is no change of parameters along $x$-axis, and since we choose the solution to have the form $\Phi_{l}^{inhom} (\bm{r},t) = \Phi_{0}^{inhom} \phi (z) \exp(-i\omega_{p_e} t + i k_{l_x} x)$,

\begin{displaymath}
\Delta \rightarrow \frac{d^2}{dz^2}-k_{l_x}^2,
\end{displaymath}
and taking into account that in homogeneous plasma
 
\begin{displaymath}
\omega_l^2 = \omega_{p_e}^2 (1 + 3 k_{l_x}^2 \lambda_D^2 + 3 k_{l_{z0}}^2 \lambda_D^2),
\end{displaymath}
where $k_{l_{z0}} = k_l \cos \psi$, we obtain

\begin{equation}
(3k_{l}^{2}\lambda _{D}^{2} \cos^2 \psi -\theta(z)\frac{z}{L}+3\lambda
_{D}^{2}\frac{d^{2}}{dz^{2}})\phi(z)=0,
\end{equation}
then, considering only the region of positive values of $z$ hereafter, that denotes the region with increasing density, we have

\begin{equation}
(\frac{d^{2}}{dz^{2}}+k_{l}^{2} \cos^2 \psi - \frac{1}{3\lambda _{D}^{2}}\frac{z}{L}%
)\phi(z)=0.
\label{eq:pre-airy}
\end{equation}

We introduce a new dimensionless variable

\begin{equation}
\tilde{z}=\left( \frac{k_l^2 L^{2}}{3 k_l^2 \lambda _{D}^{2}}\right) ^{1/3} \left(\frac{z}{L} - 3 k_l^2 \lambda_D^2 \cos^2 \psi \right),
\end{equation}
or, if we introduce a new parameter $\alpha = 3 k_l^2 \lambda_D^2 k_l L $, we may rewrite previous expression:

\begin{equation}
\tilde{z}= \alpha^{-1/3} k_l z - \alpha^{2/3} \cos^2 \psi.
\end{equation}

Eq.(\ref{eq:pre-airy}) reduces to Airy equation
\begin{equation}
\frac{d^2}{d\tilde{z}^2}\phi(\tilde{z}) - \tilde{z} \phi(\tilde{z}) = 0.
\label{eq:airy}
\end{equation}

It is convenient to write down the solution in form of Hankel functions $H^{(n)}_\nu $ $(n=1,2)$ in order to easily separate the incident and reflected wave. Within the density clump we have to regions: the conversion region, $\tilde{z}_0 < \tilde{z} <0$, where $\tilde{z}_0 = -\alpha^{2/3} \cos^2 \psi$ ($z = 0$) corresponds to a point where density starts to increase, and the region behind the reflection point $\tilde{z} > 0$. Inside the conversion region the solution is

\begin{equation}
\phi( \tilde{z}_{0} < \tilde{z} < 0) = \frac{1}{2} \sqrt{\frac{(-\tilde{z})}{3}} \left( e^{-i\pi/6} H^{(2)}_{1/3} \left(\frac{2}{3} (-\tilde{z})^{3/2} \right)  + e^{i\pi/6} H^{(1)}_{1/3} \left(\frac{2%
}{3} (-\tilde{z})^{3/2} \right) \right) = \phi_i + \phi_r,
\label{eq:airy-solution}
\end{equation}
in our case the Hankel function of second kind corresponds to incident ($i$) wave and of first kind - to reflected ($r$). After the conversion region, the wave simply damps according to solution

\begin{equation}
\phi(\tilde{z} > 0) = \frac{1}{\pi} \sqrt{\frac{\tilde{z}}{3}} K_{1/3} \left(\frac{2}{3} \tilde{z}^{3/2} \right).
\end{equation}

In order to evaluate the amplitude $\Phi_0^{inhom}$ of electrostatic potential, we will use a WKB-approximation to solve Eq.(\ref{eq:airy}). First we assume that the solution of Eq.(\ref{eq:airy}) has a form

\begin{equation}
\phi(\tilde{z})_{WKB} = \phi_0 (\tilde{z}) e^{-i k_l \Psi(\tilde{z})},
\label{eq:WKB0}
\end{equation}
where amplitude $\phi_0(\tilde{z})$ and phase $\Psi(\tilde{z})$ vary slowly with $\tilde{z}$.We subsitute this solution into Eq.(\ref{eq:airy}) and obtain ($'$ denotes $\frac{d}{d \tilde{z}}$):

\begin{equation}
\phi_0^{''} - 2ik_l \Psi^{'} \phi_0^{'} - i k_l \Psi^{''} \phi_0 - k_l^2 (\Psi^{'})^2 \phi_0 - \tilde{z} \phi_0 = 0.
\label{eq:wkb_phi}
\end{equation}

After, we divide the whole equation by $k_l^2 = 4 \pi^2 / \lambda_l^2$ and note, that according to our assumption, $\phi_0$ and $\Psi$ change noticably only on scales $l \gg \lambda_l$. For this reason we can make an estimation: $ \phi_0^{''} \sim \phi_0 / l^2 $, $\phi_0^{'} \sim \phi_0 / l$, $\Psi^{''} \sim \Psi^{'} / l$. Eq.(\ref{eq:wkb_phi}) will take a form:

\begin{equation}
\frac{\lambda_l^2}{4 \pi^2}\frac{\phi_0}{l^2} - 2i\frac{\lambda_l}{2 \pi} \frac{\phi_0}{l} \Psi^{'} - i \frac{\lambda_l}{2 \pi} \frac{\Psi^{'}}{l} \phi_0 - (\Psi^{'})^2 \phi_0 - \frac{\lambda_l^2}{4 \pi^2} \tilde{z} \phi_0 = 0.
\end{equation}

We may find an approximate solution by assigning terms of different order of $\lambda_l / l$ equal to zero:

\begin{equation}
((\Psi^{'})^2 + \frac{\lambda_l^2}{4 \pi^2} \tilde{z}) \phi_0= 0,
\label{eq:WKB1}
\end{equation}

\begin{equation}
\phi_0^{'} + \frac{\Psi^{''}}{2 \Psi^{'}} \phi_0 = 0.
\label{eq:WKB2}
\end{equation}

From Eq.(\ref{eq:WKB1}) we obtain ($\phi_0 \ne 0$):

\begin{equation}
\Psi^{'} = \pm \sqrt{- \frac{\lambda_l^2}{4 \pi^2} \tilde{z}},
\label{eq:WKB3}
\end{equation}

\begin{equation}
\Psi = \pm \sqrt{\frac{\lambda_l^2}{4 \pi^2}} \int_{\tilde{z}_0}^{\tilde{z}} \sqrt{-\tilde{z}} d\tilde{z} = \mp \frac{\lambda_l}{2 \pi} \left(\frac{2}{3} (-\tilde{z}_0)^{3/2} - \frac{2}{3} (-\tilde{z})^{3/2} \right).
\end{equation}

After we procced to the solution of Eq.(\ref{eq:WKB2}):

\begin{equation}
\phi_0 = \frac{C}{\sqrt{\Psi^{'}}},
\label{eq:WKB4}
\end{equation}
where $C$ is an integration constant. Coming back to solution (\ref{eq:WKB0}) and substituting (\ref{eq:WKB3}) into (\ref{eq:WKB4}) we obtain: 

\begin{equation}
\phi(\tilde{z})_{WKB} = \frac{C \sqrt{k_l}}{(-\tilde{z})^{1/4}} e^{\pm i \left( \frac{2}{3}(-\tilde{z}_0)^{3/2} - \frac{2}{3} (-\tilde{z})^{3/2}\right)}.
\end{equation}

The sewing will be performed at the point $\tilde{z} = \tilde{z}_0$ ($z = 0$), thus we will rewrite previous expression:

\begin{equation}
\phi(\tilde{z} = \tilde{z}_0)_{WKB} = \frac{C \sqrt{k_l}}{(-\tilde{z}_0)^{1/4}}.
\end{equation}

The complete solution for electrostatic potential in WKB-approximation will be:

\begin{equation}
\Phi_{l}^{inhom}(\tilde{z})_{WKB} = \Phi_{0}^{inhom} \frac{C \sqrt{k_l}}{(-\tilde{z})^{1/4}} e^{\pm i \left( \frac{2}{3}(-\tilde{z}_0)^{3/2} - \frac{2}{3} (-\tilde{z})^{3/2}\right) - i \omega_{p_e} t + i \zeta + ik_l \sin \psi x},
\end{equation}
where $\zeta$ is the phase difference between exact solution and solution under WKB-approximation for $\Phi_{l}^{inhom}(\tilde{z})$. Now we may procced to sewing the incident waves from homogeneous plasma with our approximate WKB - solution (we omit $e^{- i \omega_{p_e} t + ik_l \sin \psi x}$ terms, common for both waves):

\begin{equation}
\Phi^{homog}_{incid} (\tilde{z}_0) = \Phi_{0}^{homog} e^{i \eta}, 
\end{equation}

\begin{equation}
\Phi^{inhom}_{incid} (\tilde{z}_0)_{WKB} =\Phi_{0}^{inhom} \frac{C \sqrt{k_l}}{(-\tilde{z}_0)^{1/4}} e^{i\zeta}, 
\end{equation}
and putting the equal

\begin{equation}
\Phi_{0}^{homog} e^{i \eta} = \Phi_{0}^{inhom} \frac{C \sqrt{k_l}}{(-\tilde{z_0})^{1/4}} e^{i\zeta}, 
\end{equation}
we obtain $\zeta = \eta$ and

\begin{equation}
C = \frac{\Phi_{0}^{homog}}{\Phi_{0}^{inhom}}\frac{(- \tilde{z_0})^{1/4}}{\sqrt{k_l}}.
\end{equation}

Thus a solution for incident wave in WKB-approximation has a form:

\begin{equation}
\Phi^{inhom}_{incid}(\tilde{z})_{WKB} = \Phi_{0}^{homog} \frac{(- \tilde{z_0})^{1/4}}{(-\tilde{z})^{1/4}} e^{ -i \left( \frac{2}{3}(-\tilde{z}_0)^{3/2} - \frac{2}{3} (-\tilde{z})^{3/2}\right) - i \omega t + ik_l \sin \psi x + i \eta}.
\end{equation}

Now we need to figure out the phase $\eta$. In order to do this, we will use an asymptotic expansion of the exact solution for incident wave (see Eq.(\ref{eq:airy-solution})) for large value of argument:

\begin{equation}
\Phi^{inhom}_{incid}(\tilde{z})=\Phi_{0}^{inhom} \frac{(-\tilde{z})^{-1/4}}{2\sqrt{\pi} }e^{-i(\frac{2}{3}(-\tilde{z})^{3/2}-\frac{1}{4}\pi)},
\end{equation}

We set expressions for $\Phi^{inhom}_{incid}(\tilde{z})$ and $\Phi^{inhom}_{incid}(\tilde{z})_{WKB}$ equal at the point $\tilde{z} = \tilde{z}_0 = - \alpha^{2/3} \cos^2 \psi$ and obtain:

\begin{equation}
\Phi_{0}^{homog} {\cos^{1/2} \psi} e^{i \frac{2}{3} \alpha \cos^3 \psi + i \eta} = \Phi_{0}^{inhom} \frac{1}{2\sqrt{\pi} \alpha^{1/6}} e^{i\frac{\pi}{4}}.
\end{equation}

From this equation system we obtain:

\begin{equation}
\eta = \frac{\pi}{4} - \frac{2}{3}\alpha \cos^3 \psi, \qquad  \Phi_{0}^{inhom} = 2 \sqrt{\pi} \alpha^{1/6} \cos^{1/2} \psi  \Phi_{0}^{homog}.
\end{equation}

And the final expression for the electrostatic potential inside a density clump is

\begin{equation}
\Phi_{l}^{inhom}( \tilde{z}_{0} < \tilde{z} < 0) = \Phi_{0}^{homog} \sqrt{\pi} \alpha^{1/6} \cos^{1/2} \psi \sqrt{\frac{(-\tilde{z})}{3}} \left( e^{-i\pi/6} H^{(2)}_{1/3} \left(\frac{2}{3} (-\tilde{z})^{3/2} \right)  + e^{i\pi/6} H^{(1)}_{1/3} \left(\frac{2%
}{3} (-\tilde{z})^{3/2} \right) \right) e^{- i \omega t + i k_l \sin \psi x}
\end{equation}

This solution was obtained under the assumption $(-\tilde{z}_0) \gg 1$, as the WKB-approximation can be only applied in the wave zone, far from the reflection point. The criterion $(-\tilde{z}_0) \gg 1$ may be rewritten as $\alpha^{2/3} \cos^2 \psi \gg 1$, or $\alpha \gg 1$. A specific limitations on the angle of incidence $\psi$ will be discussed in Appendix \ref{app: 3}.
Corresponding electric field components can be calculated from equation $\bm{E}^{inhom} = - \nabla \Phi_{l}^{inhom} (\bm{r},t)$, and are
\begin{displaymath}
	E_{l_x}^{inhom} = -\frac{d}{d x} \Phi_{l}^{inhom} (\bm{r},t) = -i k_{l_x} \Phi_{l}^{inhom} (\bm{r},t) =
\end{displaymath}

\begin{equation}
	 = - i \sin \psi \sqrt{\pi} \alpha^{1/6} \cos^{1/2} \psi \sqrt{\frac{(-\tilde{z})}{3}} \left( e^{-i\pi/6} H^{(2)}_{1/3} \left(\frac{2}{3} (-\tilde{z})^{3/2} \right)  + e^{i\pi/6} H^{(1)}_{1/3} \left(\frac{2%
}{3} (-\tilde{z})^{3/2} \right) \right) E_{0}  e^{- i \omega_{p_e} t + ik_l \sin \psi x} = E_{l_{x_i}}^{inhom} + E_{l_{x_r}}^{inhom},
\end{equation}

\begin{displaymath}
E_{l_z}^{inhom} = -\frac{d}{d z} \Phi_{l}^{inhom} (\bm{r},t) = - \alpha^{-1/3} k_l \frac{d}{d \tilde{z}} \Phi_{l}^{inhom} (\bm{r},t) =
\end{displaymath}

\begin{eqnarray}
= -\sqrt{\pi} \alpha^{-1/6} \cos^{1/2} \psi \frac{ (-\tilde{z})}{\sqrt{3}} \left(e^{i \pi / 6} H_{2/3}^{(2)} \left( \frac{2}{3} |\tilde{z}|^{3/2} \right) + e^{-i \pi / 6} H_{2/3}^{(1)} \left(\frac{2}{3} |\tilde{z}|^{3/2} \right)\right) E_{0} e^{-i \omega_{p_e} t + i k_{l_x} x} = E_{l_{z_i}} + E_{l_{z_r}}, 
\end{eqnarray}
where we took into account $\Phi_{0}^{homog} = E_{0}^{homog} k_l^{-1}$, where $E_{0}^{homog} \equiv E_0$ is the amplitude of the electric field in a homogeneous plasma. 

The current that is excited by electron density and velocity perturbations caused by incident and reflected Langmuir wave should be written in the following form

\begin{equation}
	\bm{J}_{2 \omega_{p_e}}(\bm{r}) = -e (\delta n_{l_{i}} (\bm{r}) \bm{\delta v}_{r} (\bm{r}) + \delta n_{l_{r}} (\bm{r}) \bm{\delta v}_{i} (\bm{r})),
\end{equation}
where $\delta n_l$ and $\bm{\delta v}$ can be expressed from simplest linear relations:

\begin{equation}
	-i \omega_{p_e} \bm{\delta v}_{i, r} = -e \bm{E}_{i, r} / m_e,
\end{equation}

\begin{equation}
	-i \omega_{p_e} \delta n_{l_{i, r}} + div (n_0 \bm{\delta v}) = 0 = -i \omega_{p_e} \delta n_{l_{i, r}} + i k_{l_x} n_0 \delta v_{x_{i, r}} + n_0 \frac{d \delta v_{z_{i, r}}}{dz} ,
\end{equation}

\begin{equation}
	\bm{\delta v}_{i, r} = -i e \frac{\bm{E}_{i, r}}{\omega_{p_e} m_e},
\end{equation}

\begin{equation}
	\delta n_{l_{i, r}} = - \frac{i e n_0 k_{l_x}}{\omega_{p_e}^2 m_e} E_{l_{x_{i, r}}} - \frac{e n_0}{\omega_{p_e}^2 m_e} \frac{d E_{l_{z_{i, r}}}}{dz},
\end{equation}

The analytical expression for $J_x$ is

\begin{equation}
\begin{split}
J_{x}^{inhom} = i \frac{e^3 n_0}{\omega_l^3 m_e^2} \left( (-i k_l \sin \psi E_{x_{i}} - \frac{\partial E_{z_{i}}}{\partial z}) E_{x_{r}} + (-i k_l \sin \psi E_{x_{r}} - \frac{\partial E_{z_{r}}}{\partial z}) E_{x_{i}} \right) = 
\\
 = - \frac{1}{6} \frac{e}{m_e c} \frac{c}{v_b} \cos \psi \sin \psi \alpha^{1/3} (-\tilde{z}) \left(\sin^2 \psi +   \alpha^{-2/3} (-\tilde{z}) \right) E_{0}^2 H^{(2)}_{1/3}\left(%
\frac{2}{3}(-\tilde{z}) ^{3/2}\right) H^{(1)}_{1/3} \left(%
\frac{2}{3}(-\tilde{z}) ^{3/2}\right) e^{-2 i \omega_{p_e} t + 2 i k_{l_x} x},
\end{split}
\end{equation}
where we have used the fact that Langmuir waves are generated at local plasma frequency ($\omega_{l} \approx \omega_{p_e}$) by the electron beam under resonance condition $k_l \approx \omega_{p_e} / v_b$. 

Analytical expression for $J_z$ is
    
\begin{equation}
\begin{split}
J_{z}^{inhom} = i \frac{e^3 n_0}{\omega_{p_e}^3 m_e^2} \left( (-i k_l \sin \psi E_{x_{i}} - \frac{\partial E_{z_{i}}}{\partial z}) E_{z_{r}} + (-i k_l \sin \psi E_{x_{r}} - \frac{\partial E_{z_{r}}}{\partial z}) E_{z_{i}} \right) = 
\\
= i \frac{1}{12} \frac{e}{m_e c} \frac{c}{v_b} \cos \psi (-\tilde{z})^{3/2}  \left(\sin^2 \psi +   \alpha^{-2/3} (-\tilde{z}) \right)  E_0^2 \times 
\\
\left( H^{(2)}_{-1/3}\left(%
\frac{2}{3}(-\tilde{z}) ^{3/2}\right) H^{(1)}_{2/3} \left(%
\frac{2}{3}(-\tilde{z}) ^{3/2}\right) + H^{(1)}_{-1/3}\left(%
\frac{2}{3}(-\tilde{z}) ^{3/2}\right) H^{(2)}_{2/3} \left(%
\frac{2}{3}(-\tilde{z}) ^{3/2}\right)\right) e^{-2 i \omega_{p_e} t + 2 i k_{l_x} x}.
\end{split}
\end{equation}

\section{Emission from a single clump} \label{app: 3} %%%%%%%%%App 3
\begin{figure}[htbp]
\centering
\includegraphics[width = 0.5\textwidth]{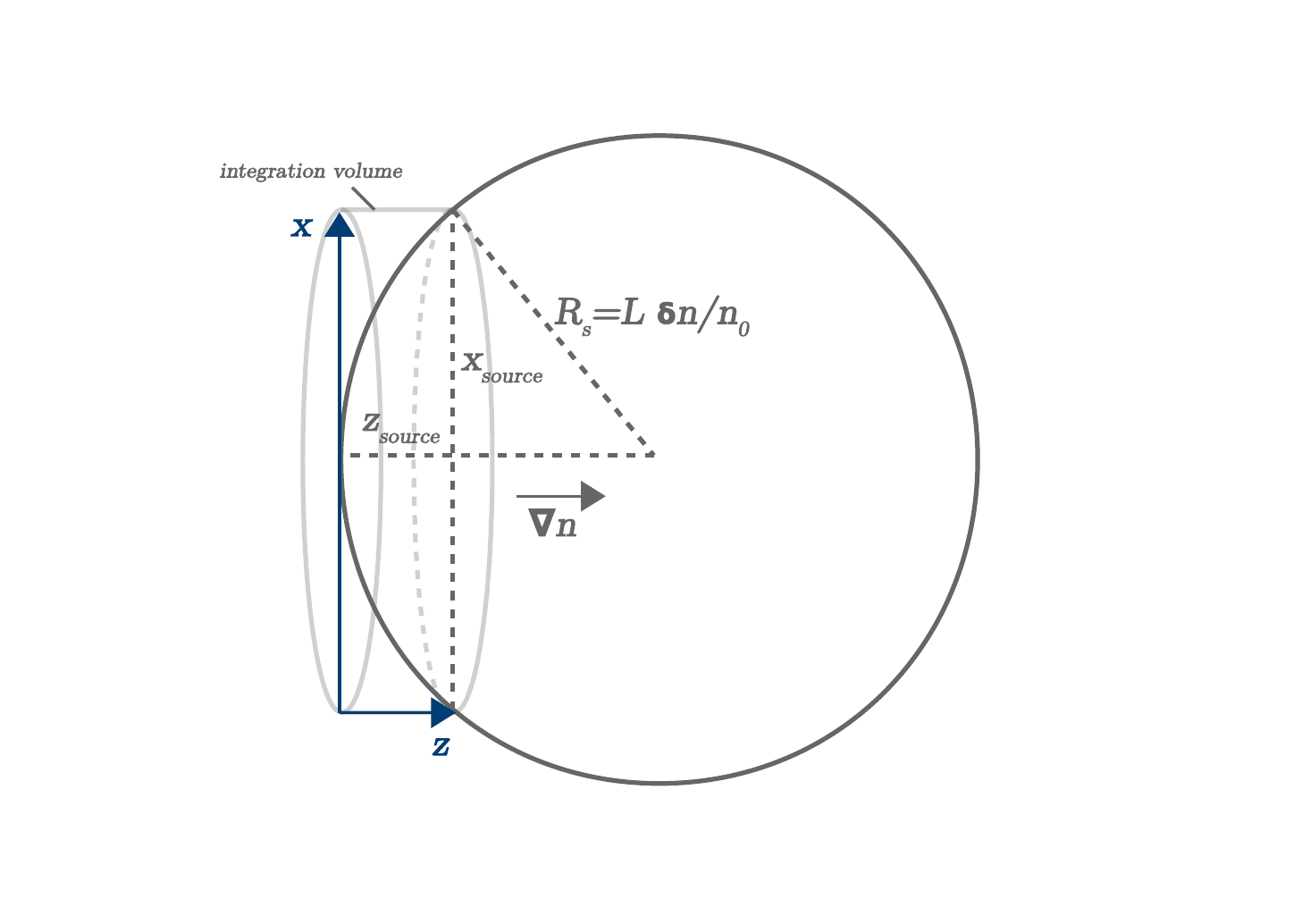}
\caption{Schematic illustration of an integration volume in the vicinity of the reflection point, where the Langmuir waves coalescence into harmonic EM emission takes place. The density clump is approximately represented as a sphere with characteristic radius $R_s = L \delta n/n_0$, while the integration volume may be roughly described as a cylinder with radius $x_{source}$ and height $z_{source}$, which are transverse and longitudinal characteristic sizes of conversion region, i.e., the region where nonlinear currents $J_{2 \omega_{p_e}}(\bm{r},t)$ are excited.}
\label{fig:int_vol}
\end{figure}

As soon as we have currents in the form
\begin{equation}
\bm{J} (\bm{r}, t^{'} ) = \bm{J} (z) e^{-2 i \omega_{p_e} t^{'} + 2 i k_{l_x} x} ,
\end{equation}
we can calculate the Lienard-Wiechert potential of the harmonic field \citep{landau2013course}, \citep{jackson2007classical}:
\begin{equation}
	\bm{A}_{2 \omega_{p_e}} (\bm{r}) e^{-2 i \omega_{p_e} t} = \frac{\sqrt{\epsilon}}{c}\int\frac{\bm{J}_{2 \omega_{p_e}}(\bm{r}, t - \frac{|\bm{R}-\bm{r}|}{c})}{|\bm{R}-\bm{r}|} d^3 r,
	\label{eq: vect_pot}
\end{equation}
and the corresponding current is
\begin{equation}
\bm{J}_{2 \omega_{p_e}}(\bm{r}, t - \frac{|\bm{R}-\bm{r}|}{c}) \approx \bm{J}_{2 \omega_{p_e}}(\bm{r}, t - \frac{R}{c} + \frac{\bm{r}\bm{n}}{c}) = \bm{J}_{2 \omega_{p_e}}(z) e^{-2 i \omega_{p_e} t + 2 i \omega_{p_e} R / c + 2 i \omega_{p_e} \bm{r}\bm{n} / c  + 2 i k_{l_x} x},
\label{eq:curr_decomp}
\end{equation}
where $2 \omega_{p_e} \bm{r}\bm{n} / c $ stands for ratio of the source size and the wavelength of an electromagnetic wave, with a factor of $4 \pi / \sqrt{3}$. We rewrite $\bm{r}\bm{n} = x\sin \theta + z \cos \theta$, and $\theta$ is the angle between wavevector of the electromagnetic emission and the $z$-axis. There are specific constrains for the value of $\psi$, for which the generation of the harmonic emission is possible. As along the $x$-axis, physical parameters are not changing according to our assumption, $x$-components of wave vectors of incident and reflected Langmuir waves should be equal $k_{l_x} = k_{l'_x}$ and the momentum conservation of three wave interaction should be applied along this axis:
\begin{equation}
k_{t_x} = k_{l_x} + k_{l'_x}, \qquad k_{t_x} = \frac{\sqrt{3} \omega_{p_e}}{c} \sin \theta, 
\qquad k_{l_x} = k_{l'_x} = \frac{\omega_{p_e}}{v_b} \sin \psi.
\end{equation}  
This way we obtain the relation between the angles $\psi$ and $\theta$ and a limitation for the angle $\psi$:
\begin{equation}
\sin \psi = \frac{\sqrt{3}}{2} \frac{v_b}{c} \sin \theta, \qquad |\psi_{\max}| \approx \frac{\sqrt{3}}{2} \frac{v_b}{c}.
\end{equation} 
Thus we may rewrite the term proportional to a source size by taking into account $k_{t} \sin \theta  = 2 k_{l} \sin \psi$:
\begin{equation}
2 i \omega_{p_e} \bm{r}\bm{n} / c = 2i \frac{\omega_{p_e}}{c}(x\sin \theta + z \cos \theta) = \frac{4i}{\sqrt{3}}\frac{v_b}{c}k_{l} x \sin \psi  + 2i \frac{\omega_{p_e}}{c} z \cos \theta,
\end{equation}
and the current will have the form
\begin{equation}
\bm{J}_{2 \omega_{p_e}}(z) e^{-2 i \omega_{p_e} t + 2 i \omega_{p_e} R / c - 2 i \omega_{p_e} \bm{r}\bm{n} / c  + 2 i k_{l_x} x} \approx \bm{J}_{2 \omega_{p_e}}(z) e^{-2 i \omega_{p_e} t + 2 i \omega_{p_e} R / c - 2i \frac{\omega_{p_e}}{c} z \cos \theta + 2 i k_{l_x} x}.
\end{equation}
We will use a decomposition for $\frac{1}{|\bm{R}-\bm{r}|}$ in spherical coordinates in order to easily separate the quadrupolar term $l = 2$:
\begin{equation}
	\frac{1}{|\bm{R}-\bm{r}|} \approx 4 \pi \sum_{l = 0}^\infty \sum_{m = -l}^l \frac{1}{2l+1} \frac{r^l}{R^{l+1}}Y^{m^*}_{l}(\theta_r, \phi_r) Y^m_l  (\theta_R, \phi_R),
	\label{eq:denom_decomp}
\end{equation}
and apply it to the Eq.(\ref{eq: vect_pot})
\begin{equation}
	\bm{A}_{2 \omega_{p_e}} (\bm{r}) e^{-2 i \omega_{p_e} t} = \frac{ 4 \pi  \sqrt{\epsilon}}{5 c} \frac{1}{R^{3}} e^{-2 i \omega_{p_e} t + 2 i \omega_{p_e} R / c}  \sum_{m = -2}^2   Y^m_2  (\theta_R, \phi_R) \int \bm{J}_{2 \omega_{p_e}}(z) e^{- 2i \frac{\omega_{p_e}}{c} z \cos \theta + 2 i k_{l_x} x} 
 Y^{m^*}_{2}(\theta_r, \phi_r) r^2 d^3 r.
\end{equation}
It is going to be convenient for us to perform integration \citep{NIST:DLMF}, \citep{prudnikov1986integrals} in a cylindrical coordinate system roughly over a cylindrical volume (see Fig.\ref{fig:int_vol}) 
\begin{equation}
\begin{split}
\bm{A}_{2 \omega_p}(\bm{r}) = \frac{4 \pi \sqrt{\epsilon}}{5 c} \frac{e^{2 i \frac{\omega_{p_e}}{c} R}}{R^3} \sum_{m = - 2}^2 Y^m_{2}(\theta_R, \phi_R) \times 
\\
\int^{z_{source}}_0 \bm{J}_{2 \omega_{p_e}}(z)e^{- 2i \frac{\omega_{p_e}}{c} z \cos \theta} dz \int^{x_{source}}_0 \rho (\rho^2 + z^2) d \rho \int_{0}^{2\pi} e^{2 i k_x \rho \cos \phi} Y^{m^*}_{2}(\tan^{-1} (\rho/z), \phi)d\phi,
\end{split}
\end{equation}
where $z_{source}$ is the distance from the start of the density gradient to the reflection point, i.e., the width of the conversion region, and it should be equal to $3 k_l^2 \lambda_D^2 L \cos^2 \psi$ (see Section \ref{sec:probl} , Appendix \ref{app: 2}). As we suppose that the characteristic radius of a spherical density clump is $R_{s} = L \delta n / n_0$, we may obtain that under condition $z_{source} \ll R_s$, which is equivalent to $3 k_l^2 \lambda_D^2 \ll \delta n / n_0$, that the transverse size of the source region is $x_{source} = \sqrt{{6 k_l^2 \lambda_D^2} / ({{\delta n} / {n_0}})} L {\delta n} / {n_0} $. 
In order to calculate the magnetic field component of the harmonic emission, we will use the following expression
\begin{equation}
\bm{H}_{t_{2 \omega_{p_e}}} = 2 i [\bm{k}_t \bm{A}_{2\omega_{p_e}}],
\end{equation}
or
\begin{eqnarray}
| H_{t_{2 \omega_{p_e}}} | = H_y = 2i(-k_{t_x} A_{z_{2 \omega_{p_e}}} + k_{t_z} A_{x_{2 \omega_{p_e}}} ).
\label{eq:H_t}
\end{eqnarray}
After calculating all the integrals and keeping only the most significant terms we obtain the $A_x$ component, which will be the most contributing :
\begin{equation}
\begin{split}
A_{x_{2 \omega_{p_e}}} = -i \frac{\sqrt{\pi \epsilon}}{8} \frac{e^{2 i R {\omega_{p_e}} / {c}- 2i \alpha \cos \theta \cos^2 \psi v_b / c + 2i \alpha \cos \theta v_b/c}}{R^3} \frac{e}{m_e c^2} \frac{c^2}{v_b^2} \frac{1}{\cos \theta} \frac{1}{k_l^5} \left( \sqrt{\frac{6 k_l^2 \lambda_D^2}{\frac{\delta n}{n_0}}} \frac{\delta n}{n_0} k_l L \right)^3 E_0^2\times \\
\left[2 \ Y^0_2  (\theta_R, \phi_R)\ J_1 \left(2 \sqrt{\frac{6 k_l^2 \lambda_D^2}{\frac{\delta n}{n_0}}} \frac{\delta n}{n_0} k_l L \sin \psi \right) + \sqrt{6} \ ( \ Y^{-2}_2  (\theta_R, \phi_R) + Y^2_2  (\theta_R, \phi_R)) J_3 \left(2 \sqrt{\frac{6 k_l^2 \lambda_D^2}{\frac{\delta n}{n_0}}} \frac{\delta n}{n_0} k_l L \sin \psi \right)\right].
\end{split}
\end{equation}
After we calculate the magnetic field component:
\begin{eqnarray}
| H_{t_{2 \omega_{p_e}}} | \approx 2 i k_{t_z} A_{x_{2 \omega_{p_e}}}.
\end{eqnarray}
And finally we can calculate the radiation energy density
\begin{eqnarray}
W_t^{inhom} = \frac{|H_{t_{2 \omega_{p_e}}}|^2}{8 \pi},
\end{eqnarray}
and integrate it by angles $\theta_R$ and $\phi_R$ using the orthonormality of spherical harmonics
\begin{equation}
\int_0^\pi \int_0^{2 \pi} Y_l^m Y_{l'}^{ m^{'^*}} d \Omega  = \delta_{l ,l'} \delta_{m, m^{'}}.
\end{equation}
The result has the form 
\begin{equation} 
\begin{split}
W_t^{inhom} = \frac{1}{8 \pi} \frac{3\pi \epsilon}{80} \frac{1}{R^6} \frac{e^2}{m_e^2 c^4} \frac{c^2}{v_b^2} \frac{1}{k_l^8} \left( \sqrt{\frac{6 k_l^2 \lambda_D^2}{\frac{\delta n}{n_0}}} \frac{\delta n}{n_0} k_l L \right)^6 \times \\
\left(4\ J_1 \left(2 \sqrt{\frac{6 k_l^2 \lambda_D^2}{\frac{\delta n}{n_0}}} \frac{\delta n}{n_0} k_l L \sin \psi \right) + 12\   J_3 \left(2 \sqrt{\frac{6 k_l^2 \lambda_D^2}{\frac{\delta n}{n_0}}} \frac{\delta n}{n_0} k_l L \sin \psi \right)\right) E_0^4.
\end{split}
\end{equation}  
We want to simplify it slightly by using limiting forms of the Bessel functions for a small value of their argument \citep{NIST:DLMF} (this approximation is always correct for rather small values of the angle $\psi$):
\begin{equation}
4  J_1^2 \left(2 \sqrt{\frac{6 k_l^2 \lambda_D^2}{\frac{\delta n}{n_0}}} \frac{\delta n}{n_0} k_l L \sin \psi \right)  + 12 J_3^2 \left(2 \sqrt{\frac{6 k_l^2 \lambda_D^2}{\frac{\delta n}{n_0}}} \frac{\delta n}{n_0} k_l L \sin \psi \right) \approx 24 \frac{\delta n}{n_0} k_l^2 L^2 k_l^2 \lambda_D^2 \sin^2 \psi.
\label{eq:psi_small}
\end{equation}
We will put the observation point $R$ at the border of a density clump to estimate the emission that is detected when it leaves the source region. Since we have implied $z_{source} \ll R_s$, this approximation used for decompositions in Eqs.(\ref{eq:curr_decomp}) and (\ref{eq:denom_decomp}) is valid and we may set $R \approx R_s = L \delta n / n_0$. Taking into account the approximate expression for Bessel functions and $k_l^2 \lambda_D^2 \approx {v_{T}^2}/{v_b^2}$ for beam-generated Langmuir waves, after few simple transformations we will obtain an expression for energy density of harmonic emission for a single density clump
\begin{equation} 
W_t^{inhom}  = 4\cdot10^2 \pi \epsilon \frac{v_{T}^6}{v_b^6} \left({\frac{k_l^2 \lambda_D^2}{\delta n / n_0}}\right)^{2} \frac{\omega_{p_e^2} L^2}{c^2} \sin^2 \psi \frac{W_{l}}{n_0 k_B T_e} W_{l}.
\label{eq: W_t}
\end{equation}

\section{Statistically averaged emission} \label{app: 4}  %%%%%%%%%App 4

In order to account for different angles of incidence, amplitudes of random density fluctuations and gradient scales, we will perform a statistical average over $\psi$, $\delta n$ and $L$ probability density functions:
\begin{equation}
\langle W_t^{inhom} \rangle_{\psi, \delta n, L} = P_{ref} \int P(\psi) P(\delta n) P(L) W_t^{inhom}(\psi, \delta n, L) \ d \psi\ d \delta n \ d L.
\end{equation}
We assume a uniform distribution over angles of incidence, $P(\psi) = \pi^{-1}$. In order to perform the integration we will remind that the value of $\psi$ is limited by $\psi_{\max} \approx {\sqrt{3}}{v_b} / 2{c} \ll 1$:
\begin{equation}
\langle W_{t}^{inhom} \rangle_{\psi}=... \frac{1}{\pi} \int\limits_{0}^{\psi_{\max}} \sin^2 \psi d \psi  \approx ... \frac{1}{\pi} \int\limits_{0}^{\psi_{\max}} \psi^2 d \psi = ... \frac{1}{3\pi} \psi_{\max}^3 \approx \frac{\sqrt{3}}{8 \pi} \frac{v_b^3}{c^3} .
\end{equation}
We shall assume that fluctuations follow a normal distribution with zero mean value and standard deviation $\langle \Delta n \rangle$:
\begin{equation}
P_{\delta n} \left( \delta n\right)=\frac{1}{\sqrt{2\pi} \langle \Delta n\rangle}\exp{\left[ -\frac{\delta n^{2}}{2 \langle \Delta n\rangle^{2}} \right]}.
\end{equation}
Then, averaging over amplitudes of density fluctuations can be performed as follows:
\begin{equation}
\langle W_{t}^{inhom} \rangle_{\delta n}=...\frac{n_0^2}{\sqrt{2\pi} \langle \Delta n\rangle}
 \int \limits_{0}^{\infty} \delta n^{-2}\exp{\left[-\frac{\delta n^{2}}{2 \langle \Delta n\rangle^{2}}\right]} d\delta n=
...-\sqrt{2} \left(\frac{\langle \Delta n \rangle}{n_0}\right)^{-2}.
\end{equation}
Let us assume that all fluctuation have the same size $P_{ref}$. In this case, the probability to find a fluctuation with density variation $\delta n$ should be equal to the probability of finding a fluctuation with gradient $L$:

\begin{equation}
P_{\delta n}\left( \delta n\right) d \delta n=P_{L}\left( L\right)dL.
\end{equation}
Since $\delta n/n_{0}= P/L$, we may write
\begin{equation}
\frac{ \partial \delta n}{ \partial L}=-\frac{n_{0}P}{L^{2}},
\end{equation}
and
\begin{equation}
P_{L}\left(L\right)dL=P_{\delta n} \left( \delta n(L) \right) \left| \frac{\partial \delta n}{\partial L}\right| dL=
\frac{1}{\sqrt{2\pi}} \frac{n_{0}}{\langle \Delta n\rangle}\frac{P}{L^{2}}\exp{\left[-\frac{P^{2}n_{0}^{2} }{2 \langle \Delta n \rangle^{2}L^{2}}\right]} \ dL,
\end{equation}
or after substitution of $P / (\langle \Delta n\rangle / n_{0}) = L_{sc}$:
\begin{equation}
P_{L}\left( L\right)dL=\frac{1}{\sqrt{2\pi}} \frac{L_{sc}}{L^{2}}\exp{\left[-\frac{L_{sc}^{2}}{2L^{2}}\right]} \ dL.
\end{equation}
Now we can perform averaging over $L$
\begin{equation}
\langle W_{t}^{inhom} \rangle_{L}=...\frac{L_{sc}}{\sqrt{2 \pi}} \int\limits_{0}^{\infty} \exp{\left[-\frac{L_{sc}^{2}}{2L^{2}}\right]}dL = ...\frac{L_{sc}^2}{2}.
\end{equation}
Putting all together, we obtain
\begin{equation}
\langle W_t^{inhom} \rangle_{\psi, \delta n, L} = {25 \sqrt{6}} \epsilon P_{ref} \frac{v_{T}^3}{c^3} \frac{v_T^3}{v_b^3} \left({\frac{k_l^2 \lambda_D^2}{\langle \Delta n \rangle / n_0}}\right)^{2} \frac{\omega_{p_e^2} L_{sc}^2}{c^2} \frac{W_{l}}{n_0 k_B T_e} W_{l}.
\label{eq:w_t}
\end{equation}

\bibliography{Harmonic_radio_emission}{}

\begin{thebibliography}{}
\expandafter\ifx\csname natexlab\endcsname\relax\def\natexlab#1{#1}\fi
\providecommand{\url}[1]{\href{#1}{#1}}
\providecommand{\dodoi}[1]{doi:~\href{http://doi.org/#1}{\nolinkurl{#1}}}
\providecommand{\doeprint}[1]{\href{http://ascl.net/#1}{\nolinkurl{http://ascl.net/#1}}}
\providecommand{\doarXiv}[1]{\href{https://arxiv.org/abs/#1}{\nolinkurl{https://arxiv.org/abs/#1}}}

\bibitem[{Bian {et~al.}(2014)Bian, Kontar, \& Ratcliffe}]{bian2014resonance}
Bian, N.~H., Kontar, E.~P., \& Ratcliffe, H. 2014, Journal of Geophysical
  Research: Space Physics, 119, 4239

\bibitem[{Brejzman \& Pekker(1978)}]{brejzman1978electromagnetic}
Brejzman, B., \& Pekker, L. 1978, Physics Letters A, 65, 121

\bibitem[{Cairns(1987{\natexlab{a}})}]{cairns1987second}
Cairns, I.~H. 1987{\natexlab{a}}, Journal of plasma physics, 38, 179

\bibitem[{Cairns(1987{\natexlab{b}})}]{cairns1987third}
---. 1987{\natexlab{b}}, Journal of plasma physics, 38, 199

\bibitem[{Celnikier {et~al.}(1987)Celnikier, Muschietti, \&
  Goldman}]{celnikier1987aspects}
Celnikier, L., Muschietti, L., \& Goldman, M. 1987, Astronomy and Astrophysics,
  181, 138

\bibitem[{Chen {et~al.}(2012)Chen, Salem, Bonnell, Mozer, \&
  Bale}]{chen2012density}
Chen, C., Salem, C., Bonnell, J., Mozer, F., \& Bale, S. 2012, Physical Review
  Letters, 109, 035001

\bibitem[{Chen {et~al.}(1984)}]{chen1984introduction}
Chen, F.~F., {et~al.} 1984, Introduction to plasma physics and controlled
  fusion, Vol.~1 (Springer)

\bibitem[{Chen {et~al.}(2018)Chen, Kontar, Yu, Yan, Huang, \&
  Tan}]{chen2018fine}
Chen, X., Kontar, E.~P., Yu, S., {et~al.} 2018, The Astrophysical Journal, 856,
  73

\bibitem[{{\relax DLMF}(2019)}]{NIST:DLMF}
{\relax DLMF}. 2019, {\it NIST Digital Library of Mathematical Functions},
  http://dlmf.nist.gov/, Release 1.0.25 of 2019-12-15.
\newblock \url{http://dlmf.nist.gov/}

\bibitem[{Drummond \& Pines(1964)}]{drummond1964nonlinear}
Drummond, W., \& Pines, D. 1964, Annals of physics, 28, 478

\bibitem[{Dulk \& Suzuki(1980)}]{dulk1980position}
Dulk, G., \& Suzuki, S. 1980, Astronomy and Astrophysics, 88, 203

\bibitem[{Dulk {et~al.}(1998)Dulk, Leblanc, Robinson, Bougeret, \&
  Lin}]{dulk1998electron}
Dulk, G.~A., Leblanc, Y., Robinson, P.~A., Bougeret, J.-L., \& Lin, R.~P. 1998,
  Journal of Geophysical Research: Space Physics, 103, 17223

\bibitem[{Ergun {et~al.}(1998)Ergun, Larson, Lin, McFadden, Carlson, Anderson,
  Muschietti, McCarthy, Parks, Reme, {et~al.}}]{ergun1998wind}
Ergun, R., Larson, D., Lin, R., {et~al.} 1998, The Astrophysical Journal, 503,
  435

\bibitem[{Ergun {et~al.}(2008)Ergun, Malaspina, Cairns, Goldman, Newman,
  Robinson, Eriksson, Bougeret, Briand, Bale, {et~al.}}]{ergun2008eigenmode}
Ergun, R., Malaspina, D., Cairns, I.~H., {et~al.} 2008, Physical review
  letters, 101, 051101

\bibitem[{Erokhin {et~al.}(1974)Erokhin, Moiseev, \&
  Mukhin}]{erokhin1974theory}
Erokhin, N., Moiseev, S., \& Mukhin, V. 1974, Nuclear Fusion, 14, 333

\bibitem[{Galeev \& Krasnoselskikh(1976)}]{galeev1976strong}
Galeev, A., \& Krasnoselskikh, V. 1976, ZhETF Pisma Redaktsiiu, 24, 558

\bibitem[{Ginzburg \& Zheleznyakov(1958)}]{ginzburg1958possible}
Ginzburg, V., \& Zheleznyakov, V. 1958, Soviet Astron. AJ, 2, 235

\bibitem[{Goldman {et~al.}(1980)Goldman, Reiter, \&
  Nicholson}]{goldman1980radiation}
Goldman, M.~V., Reiter, G.~F., \& Nicholson, D.~R. 1980, The Physics of Fluids,
  23, 388

\bibitem[{Goldstein {et~al.}(1995)Goldstein, Roberts, \&
  Matthaeus}]{goldstein1995magnetohydrodynamic}
Goldstein, M.~L., Roberts, D.~A., \& Matthaeus, W. 1995, Annual review of
  astronomy and astrophysics, 33, 283

\bibitem[{Gurnett \& Frank(1978)}]{gurnett1978ion}
Gurnett, D.~A., \& Frank, L.~A. 1978, Journal of Geophysical Research: Space
  Physics, 83, 58

\bibitem[{Harding {et~al.}(2020)Harding, Cairns, \&
  Melrose}]{harding2020electron}
Harding, J.~C., Cairns, I.~H., \& Melrose, D.~B. 2020, Physics of Plasmas, 27,
  020702

\bibitem[{Hinkel-Lipsker {et~al.}(1992)Hinkel-Lipsker, Fried, \&
  Morales}]{hinkel1992analytic}
Hinkel-Lipsker, D., Fried, B., \& Morales, G. 1992, Physics of Fluids B: Plasma
  Physics, 4, 559

\bibitem[{Jackson(2007)}]{jackson2007classical}
Jackson, J.~D. 2007, Classical electrodynamics (John Wiley \& Sons)

\bibitem[{Kellogg(1980)}]{kellogg1980fundamental}
Kellogg, P. 1980, The Astrophysical Journal, 236, 696

\bibitem[{Kellogg {et~al.}(1999)Kellogg, Goetz, Monson, \&
  Bale}]{kellogg1999langmuir}
Kellogg, P., Goetz, K., Monson, S., \& Bale, S. 1999, Journal of Geophysical
  Research: Space Physics, 104, 17069

\bibitem[{Kellogg \& Horbury(2005)}]{kellogg2005rapid}
Kellogg, P.~J., \& Horbury, T. 2005, Annales Geophysicae, 23, 3765

\bibitem[{Kim {et~al.}(2013)Kim, Cairns, \& Johnson}]{kim2013linear}
Kim, E.-H., Cairns, I.~H., \& Johnson, J.~R. 2013, Physics of Plasmas, 20,
  122103

\bibitem[{Kim {et~al.}(2007)Kim, Cairns, \& Robinson}]{kim2007extraordinary}
Kim, E.-H., Cairns, I.~H., \& Robinson, P.~A. 2007, Physical review letters,
  99, 015003

\bibitem[{Kim {et~al.}(2008)Kim, Cairns, \& Robinson}]{kim2008mode}
---. 2008, Physics of Plasmas, 15, 102110

\bibitem[{Kim {et~al.}(2009)Kim, Johnson, Cairns, \& Lee}]{kim2009waves}
Kim, E.-H., Johnson, J.~R., Cairns, I.~H., \& Lee, D.-H. 2009in , American
  Institute of Physics, 13--20

\bibitem[{Kontar {et~al.}(2019)Kontar, Chen, Chrysaphi, Jeffrey, Emslie,
  Krupar, Maksimovic, Gordovskyy, \& Browning}]{kontar2019anisotropic}
Kontar, E.~P., Chen, X., Chrysaphi, N., {et~al.} 2019, The Astrophysical
  Journal, 884, 122

\bibitem[{Krafft \& Volokitin(2014)}]{krafft2014hamiltonian}
Krafft, C., \& Volokitin, A. 2014, The European Physical Journal D, 68, 370

\bibitem[{Krafft {et~al.}(2013)Krafft, Volokitin, \&
  Krasnoselskikh}]{krafft2013interaction}
Krafft, C., Volokitin, A., \& Krasnoselskikh, V. 2013, The Astrophysical
  Journal, 778, 111

\bibitem[{Krasnoselskikh {et~al.}(2011)Krasnoselskikh, Dudok~de Wit, \&
  Bale}]{krasnoselskikh2011determining}
Krasnoselskikh, V., Dudok~de Wit, T., \& Bale, S. 2011, Annales Geophysicae,
  29, 613, \dodoi{10.5194/angeo-29-613-2011}

\bibitem[{Krasnoselskikh {et~al.}(2007)Krasnoselskikh, Lobzin, Musatenko,
  Soucek, Pickett, \& Cairns}]{krasnoselskikh2007beam}
Krasnoselskikh, V., Lobzin, V., Musatenko, K., {et~al.} 2007, Journal of
  Geophysical Research: Space Physics, 112

\bibitem[{Krasnoselskikh {et~al.}(2019)Krasnoselskikh, Voshchepynets, \&
  Maksimovic}]{krasnoselskikh2019efficiency}
Krasnoselskikh, V., Voshchepynets, A., \& Maksimovic, M. 2019, The
  Astrophysical Journal, 879, 51

\bibitem[{{Krucker} {et~al.}(2009){Krucker}, {Oakley}, \&
  {Lin}}]{Krucker_Oakley_Lin_2009}
{Krucker}, S., {Oakley}, P.~H., \& {Lin}, R.~P. 2009, The Astrophysical
  Journal, 691, 806, \dodoi{10.1088/0004-637X/691/1/806}

\bibitem[{Krupar {et~al.}(2020)Krupar, Szabo, Maksimovic, Kruparova, Kontar,
  Balmaceda, Bonnin, Bale, Pulupa, Malaspina, {et~al.}}]{krupar2020density}
Krupar, V., Szabo, A., Maksimovic, M., {et~al.} 2020, The Astrophysical Journal
  Supplement Series, 246, 57

\bibitem[{Landau \& Lifshitz(2013)}]{landau2013course}
Landau, L.~D., \& Lifshitz, E.~M. 2013, Course of theoretical physics
  (Elsevier)

\bibitem[{Lin {et~al.}(1986)Lin, Levedahl, Lotko, Gurnett, \&
  Scarf}]{lin1986evidence}
Lin, R., Levedahl, W., Lotko, W., Gurnett, D., \& Scarf, F. 1986, The
  Astrophysical Journal, 308, 954

\bibitem[{Lin {et~al.}(1981)Lin, Potter, Gurnett, \& Scarf}]{lin1981energetic}
Lin, R., Potter, D., Gurnett, D., \& Scarf, F. 1981, The Astrophysical Journal,
  251, 364

\bibitem[{Malaspina \& Ergun(2008)}]{malaspina2008observations}
Malaspina, D., \& Ergun, R. 2008, Journal of Geophysical Research: Space
  Physics, 113

\bibitem[{Malaspina {et~al.}(2012)Malaspina, Cairns, \&
  Ergun}]{malaspina2012antenna}
Malaspina, D.~M., Cairns, I.~H., \& Ergun, R.~E. 2012, The Astrophysical
  Journal, 755, 45

\bibitem[{Mann {et~al.}(2018)Mann, Breitling, Vocks, Aurass, Steinmetz,
  Strassmeier, Bisi, Fallows, Gallagher, Kerdraon, {et~al.}}]{mann2018tracking}
Mann, G., Breitling, F., Vocks, C., {et~al.} 2018, Astronomy \& Astrophysics,
  611, A57

\bibitem[{Melrose(1980{\natexlab{a}})}]{melrose1980emission}
Melrose, D. 1980{\natexlab{a}}, Space Science Reviews, 26, 3

\bibitem[{Melrose(1980{\natexlab{b}})}]{melrose1980plasma}
---. 1980{\natexlab{b}}, New York, Gordon and Breach Science Publishers, 1980.
  430 p

\bibitem[{Melrose(1987)}]{melrose1987plasma}
---. 1987, in Particle Acceleration and Trapping in Solar Flares (Springer),
  89--101

\bibitem[{Meyer-Vernet \& Issautier(1998)}]{meyer1998electron}
Meyer-Vernet, N., \& Issautier, K. 1998, Journal of Geophysical Research: Space
  Physics, 103, 29705

\bibitem[{Mj{\o}lhus(1983)}]{mjolhus1983linear}
Mj{\o}lhus, E. 1983, Journal of plasma physics, 30, 179

\bibitem[{Mj{\o}lhus(1990)}]{mjolhus1990linear}
---. 1990, Radio science, 25, 1321

\bibitem[{Neugebauer(1975)}]{neugebauer1975enhancement}
Neugebauer, M. 1975, Journal of Geophysical Research, 80, 998

\bibitem[{Nishikawa \& Ryutov(1976)}]{nishikawa1976relaxation}
Nishikawa, K., \& Ryutov, D. 1976, Journal of the Physical Society of Japan,
  41, 1757

\bibitem[{Papadopoulos \& Freund(1978)}]{papadopoulos1978solitons}
Papadopoulos, K., \& Freund, H. 1978, Geophysical Research Letters, 5, 881

\bibitem[{Prudnikov {et~al.}(1986)Prudnikov, Brychkov, \&
  Marichev}]{prudnikov1986integrals}
Prudnikov, A.~P., Brychkov, Y.~A., \& Marichev, O.~I. 1986, Integrals and
  series: special functions, Vol.~2 (CRC Press)

\bibitem[{{Ratcliffe} {et~al.}(2012){Ratcliffe}, {Bian}, \&
  {Kontar}}]{Ratcliffe_Kontar_2012}
{Ratcliffe}, H., {Bian}, N.~H., \& {Kontar}, E.~P. 2012, The Astrophysical
  Journal, 761, 176, \dodoi{10.1088/0004-637X/761/2/176}

\bibitem[{Reid \& Kontar(2010)}]{reid2010solar}
Reid, H.~A., \& Kontar, E.~P. 2010, The Astrophysical Journal, 721, 864

\bibitem[{Reid \& Ratcliffe(2014)}]{reid2014review}
Reid, H. A.~S., \& Ratcliffe, H. 2014, Research in Astronomy and Astrophysics,
  14, 773

\bibitem[{Robinson \& Cairns(1998)}]{robinson1998fundamental}
Robinson, P., \& Cairns, I. 1998, Solar Physics, 181, 363

\bibitem[{Schleyer {et~al.}(2013)Schleyer, Cairns, \& Kim}]{schleyer2013linear}
Schleyer, F., Cairns, I.~H., \& Kim, E.-H. 2013, Physics of Plasmas, 20, 032101

\bibitem[{Schleyer {et~al.}(2014)Schleyer, Cairns, \& Kim}]{schleyer2014linear}
---. 2014, Journal of Geophysical Research: Space Physics, 119, 3392

\bibitem[{Shaikh \& Zank(2010)}]{shaikh2010turbulent}
Shaikh, D., \& Zank, G. 2010, Monthly Notices of the Royal Astronomical
  Society, 402, 362

\bibitem[{Sturrock(1964)}]{sturrock1964type}
Sturrock, P. 1964, NASA Special Publication, 50, 357

\bibitem[{Thejappa {et~al.}(2007)Thejappa, MacDowall, \&
  Kaiser}]{thejappa2007monte}
Thejappa, G., MacDowall, R., \& Kaiser, M. 2007, The Astrophysical Journal,
  671, 894

\bibitem[{Tsytovich(2012)}]{tsytovich2012nonlinear}
Tsytovich, V. 2012, Nonlinear effects in plasma (Springer Science \& Business
  Media)

\bibitem[{Vedenov {et~al.}(1962)Vedenov, Velikhov, \&
  Sagdeev}]{vedenov1962quasi}
Vedenov, A.~A., Velikhov, E.~P., \& Sagdeev, R.~Z. 1962, Quasi-linear theory of
  plasma oscillations, Tech. rep., Kurchatov Inst. of Atomic Energy, Moscow

\bibitem[{{Volokitin} \& {Krafft}(2016)}]{Volokitin_Krafft_2016}
{Volokitin}, A.~S., \& {Krafft}, C. 2016, The Astrophysical Journal, 833, 166,
  \dodoi{10.3847/1538-4357/833/2/166}

\bibitem[{{Volokitin} \& {Krafft}(2018)}]{Volokitin_Krafft_2018}
---. 2018, The Astrophysical Journal, 868, 104,
  \dodoi{10.3847/1538-4357/aae7cc}

\bibitem[{{Volokitin} \& {Krafft}(2020)}]{Volokitin_Krafft_2020}
---. 2020, The Astrophysical Journal Lettersl, 893, L47,
  \dodoi{10.3847/2041-8213/ab74de}

\bibitem[{{Voshchepynets} \&
  {Krasnoselskikh}(2013)}]{Voshchepynets_Krasnoselskikh_2013}
{Voshchepynets}, A., \& {Krasnoselskikh}, V. 2013, Annales Geophysicae, 31,
  1379, \dodoi{10.5194/angeo-31-1379-2013}

\bibitem[{Voshchepynets \&
  Krasnoselskikh(2015)}]{voshchepynets2015probabilistic2}
Voshchepynets, A., \& Krasnoselskikh, V. 2015, Journal of Geophysical Research:
  Space Physics, 120, 10

\bibitem[{Voshchepynets {et~al.}(2015)Voshchepynets, Krasnoselskikh, Artemyev,
  \& Volokitin}]{voshchepynets2015probabilistic}
Voshchepynets, A., Krasnoselskikh, V., Artemyev, A., \& Volokitin, A. 2015, The
  Astrophysical Journal, 807, 38

\bibitem[{Voshchepynets {et~al.}(2017)Voshchepynets, Volokitin, Krasnoselskikh,
  \& Krafft}]{voshchepynets2017statistics}
Voshchepynets, A., Volokitin, A., Krasnoselskikh, V., \& Krafft, C. 2017,
  Journal of Geophysical Research: Space Physics, 122, 3915

\bibitem[{Wasow(2018)}]{wasow2018asymptotic}
Wasow, W. 2018, Asymptotic expansions for ordinary differential equations
  (Courier Dover Publications)

\bibitem[{Willes {et~al.}(1996)Willes, Robinson, \& Melrose}]{willes1996second}
Willes, A., Robinson, P., \& Melrose, D. 1996, Physics of Plasmas, 3, 149

\bibitem[{Zakharov(1972)}]{zakharov1972collapse}
Zakharov, V.~E. 1972, Sov. Phys. JETP, 35, 908

\bibitem[{Zheleznyakov \& Zaitsev(1970)}]{zheleznyakov1970theory}
Zheleznyakov, V., \& Zaitsev, V. 1970, Soviet Astronomy, 14, 250

\bibitem[{Ziebell {et~al.}(2008)Ziebell, Gaelzer, Pavan, \&
  Yoon}]{ziebell2008two}
Ziebell, L., Gaelzer, R., Pavan, J., \& Yoon, P. 2008, Plasma Physics and
  Controlled Fusion, 50, 085011

\bibitem[{Ziebell {et~al.}(2011)Ziebell, Yoon, Pavan, \&
  Gaelzer}]{ziebell2011two}
Ziebell, L., Yoon, P., Pavan, J., \& Gaelzer, R. 2011, Plasma Physics and
  Controlled Fusion, 53, 085004

\end{thebibliography}
\bibliographystyle{aasjournal}

%% This command is needed to show the entire author+affiliation list when
%% the collaboration and author truncation commands are used.  It has to
%% go at the end of the manuscript.
%\allauthors

%% Include this line if you are using the \added, \replaced, \deleted
%% commands to see a summary list of all changes at the end of the article.
%\listofchanges

\end{document}